\documentclass{aa}  

\usepackage{graphicx}
\usepackage{xcolor}
\usepackage{amsmath}
\usepackage{natbib}
\usepackage{caption}
\usepackage{subcaption}
\usepackage{tikz}
\usetikzlibrary{decorations.pathmorphing}
\usepackage{soul}
\usepackage{txfonts}

\begin{document} 
\newcommand{\rh}[1]{{\color{red} #1}}
\newcommand{\sa}[1]{{\color{blue} #1}}
\newcommand{\ez}[1]{{\color{purple} #1}}
   \title{Cosmic ray transport and acceleration in an evolving shock landscape}


   \author{S.~Aerdker
          \inst{1}\inst{2}
          \and
          R.~Habegger
          \inst{3,4}
          \and 
          L.~Merten
          \inst{1}\inst{2}
          \and
          E.~Zweibel
          \inst{3,4}
          \and 
          J.~Becker Tjus
          \inst{1}\inst{2}\inst{5}
          }

  \institute{ Ruhr-Universit\"at Bochum, Fakult\"at f\"ur Physik und Astronomie, Institut f\"ur Theoretische Physik IV, Universit\"atsstra\ss e 150, 44780 Bochum, Germany \and Ruhr Astroparticle and Plasma Physics Center (RAPP Center), Bochum, Germany \and Department of Astronomy, University of Wisconsin–Madison, 475 N. Charter Street, Madison, WI 53706, USA \and Department of Physics, University of Wisconsin–Madison, 150 University Avenue, Madison, WI 53706-1390, USA \and Chalmers University of Technology, Department of Space, Earth and Environment,  412 96 Gothenburg, Sweden}

   \date{\today}

 
  \abstract
   {Cosmic rays are detected from $10^{9}$~eV up to $10^{20}$~eV, with two distinct changes in the spectrum at $\sim 10^{15}$~eV and at $10^{18.5}$~eV. Below the first break (the so-called cosmic-ray knee), sources of acceleration are believed to be located in the Milky Way. Above the second break (the so-called cosmic-ray ankle), the population is expected to be extragalactic. In between, there is the need for a third population of sources being able to accelerate particles to extreme energies. The Galactic wind and its termination shock have been proposed to fill the gap, but another possibility is large-scale shock structures from local sources in the Milky Way.}
   {In this paper, we investigate CR transport in a time-dependent landscape of shocks in the Galactic halo. These shocks could result from local outbursts, e.g.\ starforming regions and superbubbles. CRs re-accelerated at such shocks can reach energies above the knee. Since the shocks are closer to the Galaxy than a termination shock and CRs escape downstream, they can propagate back more easily. With such outbursts happening frequently, shocks will interact. This interaction could adjust the CR spectrum, particularly for the particles that are able to be accelerated at two shocks simultaneously.} 
   {The transport and acceleration of CRs at the shock is modeled by Stochastic Differential Equations (SDEs) within the public CR propagation framework CRPropa. We developed extensions for time-dependent wind profiles and for the first time connected the code to hydrodynamic simulations, which were run with the public Athena++ code. }
   {We find that, depending on the concrete realization of the diffusion tensor, a significant fraction of CRs can make it back to the Galaxy. These could contribute to the observed spectrum around and above the CR knee ($E \gtrsim 10\,\mathrm{PeV}$). In contrast to simplified models, a simple power-law does not describe the energy spectra well. Instead, for single shocks, we find a flat spectrum ($E^{-2}$) at low energies, which steepens gradually until it reaches an exponential decline. When shocks collide, the energy spectra transiently become harder than $E^{-2}$ at high energies.
    }
   {}

   \maketitle

\section{Introduction}
\label{sec:intro}


Cosmic rays (CRs) have been studied for decades and the CR spectrum and composition has been constrained across more than 10 decades in energy, but the origins of CRs are still unclear. While there are good arguments that CRs below the cosmic-ray knee come from local sources in the Milky Way and CRs above the CR ankle ($10^{18.5}$~eV) originate from extragalactic sources, the mechanisms for the energy spectrum between the two kinks is still under debate \citep{BeckerTjusMerten2020}. 
In this energy range, the transition from Galactic to extragalactic sources must occur and is also referred to as transition region. It has been discussed previously that the observed spectrum cannot simply be explained by one declining and one rising component, but that an additional source population - previously referred to as "component B" - must exist to fill the gap \citep{hillas-2005}. Note that the transition between ballistic and diffusive transport is also expected in this energy range between the knee and the ankle for the Galactic magnetic field.

It has been argued by several groups \citep{JokipiiMorfill87, ThoudamEA2016, BustardEtAl2017, Merten-etal-2018, MukhopadhyayEtAl2023} that a Galactic wind termination shock (GWTS) could (re-)accelerate Galactic CRs to the transition region. Previous models often assumed a stationary wind profile (see e.g.~\cite{JokipiiMorfill87}, \cite{Merten-etal-2018}, \cite{ThoudamEA2016}). This is a very optimistic assumption as the relevant time scales --- shock lifetime, time for the wind to reach the termination shock, and CR acceleration time scale --- are on the same order in magnitude. Although CRs might be accelerated to the desired energies it is very hard for them to propagate back into the Galaxy because the termination shock is far outside of the Galactic disk ($R_\mathrm{shock}\gtrsim 200\,\mathrm{kpc}$). Additionally, they have to escape upstream, propagate against the wind and experience adiabatic energy loss in the expanding wind.

In particular, for starburst galaxies, evidence for strong supersonic winds can be found \citep{LopezEA2017, VeilleuxEA2005}. And, it is believed that CR feedback plays an important role in Galaxy evolution, star formation and driving of galactic winds. Star formation and outflows may be linked through the Parker instability: Non-thermal CR pressure --- e.g.\ in regions of strong star formation --- perturbs the magnetic field lines tangential to the Galactic plane, forming magnetic troughs that gas can flow in. This can launch local galactic winds or fountains (see e.g.\ \citet{Ruszkowski-Pfrommer-2023} and references therein). 

In addition to the GWTS, many other source classes, such as microquasars (see e.g. \cite{Abaroa-et-al-2024}) and superbubbles \citep{Bykov-Fleishman-1992, Vieu-etal-2022, Vieu-etal-2022-II}, have been proposed to contribute to the CR flux around and above $\sim\,\mathrm{PeV}$ energies. In this paper, we investigate particle acceleration at non-stationary shocks moving out of the Galaxy. This has several advantages compared to the GWTS scenario: Such shocks can be closer to the Galaxy, so that the distance re-accelerated CRs have to cover is way less than $200\,\mathrm{kpc}$. In a blast wave scenario, the upstream region is outside of the shocked sphere, thus, re-accelerated CRs have to escape downstream to be observed at Earth. On the other hand, the downstream frame now moves with the shock speed away from the Galaxy. However, if the shock slows down over time, e.g.\ due to spherical expansion, this constraint becomes less severe. This makes the escape from the acceleration region and propagation back to the Galaxy more likely than at a GWTS. Furthermore, CRs spend less time in the expanding wind --- or the wind slows itself down --- which decreases adiabatic energy losses. 

Local, non-stationary shock structures can result from perturbations in the Galaxy. Examples are strong supernova (SN) or superbubble (SB) explosions driving large-scale outflows \citep{Habegger-2023}. While e.g.\ the forward shock of an SB might have negligible speed within the Galactic disk \citep{Castor-etal-1975, Weaver-etal-1977, Vieu-etal-2022-II}, a shock can reach considerably high speeds when small perturbations happen close to the edge of the Galaxy and trigger the Parker instability \citep{Dorfi-etal-2012, Dorfi-etal-2019}. Therefore we focus on those regions close to the edge of the Galaxy. %

Observations of magnetized structures in the Galactic halo seem to be connected to starforming regions in the Galactic disk \citep{Zhang-etal-2024}. They could be sustained from such local outflows. Furthermore, evidence of the presence of relativistic electrons in the Galactic halo has been reported \citep{Zhang-etal-2024}, which provides a hint on acceleration sites high above the Galactic disk. Electrons are not expected to propagate over large distances due to their small diffusion coefficient and short loss time scales.

In this paper, we take a closer look at such a scenario by studying CR acceleration at a spherical shock similar to the Sedov-Taylor blast wave which is launched with tens to hundreds of a typical supernova's energy and moves into a medium with a decreasing density profile. We derive acceleration times and estimates of the maximal energy reached at the shock. Further, we obtain time-dependent energy spectra at the shock and the Galactic boundary from simulations. These spectra could potentially be observed at Earth. 

With several of those perturbations happening frequently in the Galaxy, those structures build a shock landscape in the Galactic halo and contribute to a global wind structure \citep{Dorfi-etal-2012, Dorfi-etal-2019}. Additionally, if several outbursts occur near each other in both space and time, they will likely interact when they expand outwards and might run into each other. In this work we further study particle acceleration at two consecutive shocks that eventually merge. 
Similar considerations have been made for converging stellar winds and supernova in starforming regions in a steady-state \citep{Bykov-etal-2013} and time-dependent way \citep{Vieu-etal-2020}. Or, for the interaction of shocks driven by coronal mass ejections in the solar wind \citep{Wang-etal-2017, Wang-etal-2019}. In this work, we not only obtain the time-dependent energy spectra at the individual colliding shocks but also at the single shock that develops after merging. The wind profile after the collision is given by the flow profile of the Sod shock tube problem \citep{1978Sod}. We further investigate the probability of re-accelerated CRs to diffuse back to the Galaxy.

The transport and acceleration of colliding shocks is modeled in one dimension and planar shocks are considered. While the scenario is idealized, it is a step toward understanding how the complex shock interactions --- which drive the Galactic wind (and the termination shock) --- could affect CR acceleration. 

In order to obtain the spectral slope and model the propagation of CRs from the moving shocks back to the Galaxy we use a modified version of the publicly available cosmic ray transport framework CRPropa (\citet{CRPropa3.2, MertenEA2017}, hereafter M17). The transport equations of cosmic rays are solved through stochastic differential equations (SDEs). The framework is tested and well suited for studies of CR acceleration in different scenarios (\citet{Aerdker-etal-2024}, hereafter A24; \citet{Merten-Aerdker-2024}, hereafter MA24). Where no simple analytical description of the wind profiles is available, we use Athena++ \citep{Athena++} for hydrodynamic simulations. 

The paper is organized as follows: In section \ref{sec:implementation} the basics of the implementation of CRPropa based SDE solver are explained. In addition, we discuss in detail all new extensions, e.g., specialized adaptive steps and time-dependent wind profiles, that have been developed to accurately model the scenarios discussed above. Furthermore, all new code parts are validated. The updated software framework is applied to the individual time dependent spherical shocks propagating into the Galactic halo in section \ref{sec:sphericalwind}. Section \ref{sec:twoshocks} includes the modeling of the two shocks and we discuss how the wind profile of the colliding shocks is constructed. The paper concludes with a summary and discussion of the results in section \ref{sec:discussion}.

\section{Transport description}
\label{sec:implementation}
In order to investigate transport and acceleration of CRs in local and time-dependent structures, we use a modified version of CRPropa 3.2 (MA24). The transport equation is solved by rewriting it into the corresponding set of stochastic differential equations (SDEs). For details on the SDE solver that comes with the publicly available version we refer to (M17)). We give a brief overview on the algorithm, but otherwise refer to (M17, MA24). 

\subsection{From transport equations to SDEs}

The transport of CRs is modeled by the partial differential equation
\begin{equation}
     \frac{\partial f}{\partial t} = \nabla \cdot \left(\hat{\kappa} \nabla f \right) - \mathbf{u} \cdot \nabla f + \frac{p}{3} \nabla \cdot \mathbf{u} \frac{\partial f}{\partial p} + S(\mathbf{x}, p, t) \label{eq:transport_dist} \quad, 
\end{equation}
which describes the time evolution of the isotropic part of the distribution function $f(\vec{x}, p, t)$ with respect to advection $\vec{u}(\vec{x},t)$, diffusion, given by a spatial diffusion tensor $\hat{\kappa}(\vec{x}, p)$, adiabatic cooling/heating through the $\nabla \cdot \mathbf{u}$ term,  and sources/sinks $S(\vec{x}, p, t)$. 

In order to rewrite it into a set of SDEs, it has to be transformed to a Fokker-Planck equation (MA24)). The easiest way to do this is to consider  the number density $\mathcal{N} = fp^2$ instead of the distribution function:
\begin{equation}
   \frac{\partial \mathcal{N}}{\partial t} = \frac{1}{2} \nabla^2 (2\hat{\kappa}\mathcal{N}) - \nabla \left[\left(\nabla\hat{\kappa} + \mathbf{u} \right) \mathcal{N} \right] 
    - \frac{\partial}{\partial p} \left[ - \frac{p}{3}(\nabla \cdot \mathbf{u})\mathcal{N} \right] + \tilde{S} \quad . \label{eq:transport_number}
\end{equation}
%

%
%

When rewriting eq.~(\ref{eq:transport_number}) into SDEs, it splits up into one equation describing the spatial transport and one describing changes in momentum, which is 
similar to the operator splitting \textit{ansatz} often used to solve the FP equation. The spatial evolution of a phase-space element or \emph{pseudo-particle} is given by
\begin{equation}
    \label{eq:SDE}
    \mathrm{d}\vec{x} = \left( \vec{u}(\vec{x},t) + \nabla \cdot \hat{\kappa}(\vec{x}, p) \right) \mathrm{d}t + \sqrt{2 \hat{\kappa}(\vec{x}, p)} \mathrm{d}\mathbf{W}_t\quad,
\end{equation}
with the Wiener process $\mathrm{d}\mathbf{W}_t$ as a stochastic driver for the diffusive part. Equation (\ref{eq:SDE}) is solved with an Euler-Maruyama scheme with time step $\Delta t$, where the Wiener process is approximated by $\mathrm{d}\mathbf{W}_{t + 1} - \mathrm{d}\mathbf{W}_{t}  = \sqrt{\Delta t} \,\vec{\eta}_x$. The random numbers $\vec{\eta}_x$ are drawn from a normal distribution with mean value of zero. Details are presented in e.g.~M17 and MA24. For quickly changing advection, more advanced schemes like the predictor-corrector scheme might be crucial to obtain correct results. However, in our applications we find that the Euler-Maruyama scheme produces the expected results, when the time step is chosen correctly. 

The diffusion tensor $\hat{\kappa}$ is usually defined in local magnetic field coordinates, where the diagonal entries describe transport parallel and perpendicular to the magnetic field line. Off-diagonal entries describe drifts due to the curvature and gradient of the magnetic field. In this work, we only consider diffusion parallel and perpendicular to the magnetic field. For details on the integration of the magnetic field line in each simulation step, see M17 and MA24.

The time evolution in momentum is given by an ordinary differential equation since diffusion in momentum space --- also known as second order Fermi acceleration --- is not considered (it is expected to be ${\mathcal{O}}(v_A^2/c^2)$ relative to spatial diffusion). In the diffusive or macroscopic picture, energy gain at the shock comes from the compression of the background flow. The differential equation
\begin{equation}
    \label{eq:ODE}
    \mathrm{d}p = - \frac{p}{3} \nabla \cdot \vec{u}(\vec{x}, t) \mathrm{d}t\quad,
\end{equation}
is also integrated with an Euler scheme analogously to the SDE. 

Note that, equations (\ref{eq:SDE}) and (\ref{eq:ODE}) do not describe trajectories and energy gain of physical particles but of samples of the differential number density $\mathcal{N} = fp^2$. The distribution function or number density is then obtained from an ensemble average over a large number of pseudo-particle trajectories. We refer to MA24 for details on creating histograms, e.g.~of the energy spectrum and a detailed analysis of the uncertainty given by the Monte Carlo error $X/\sqrt{N}$ of the property $X$ and the number of independent\footnote{Pseudo-particles can depend on each other when the \texttt{CandidateSplitting} module is used to increase statistics at high energies.} pseudo-particles $N$, e.g.~in an energy bin $E_i$.

The SDE approach is in principle independent on specific boundary conditions or source terms. Thus, as long as the phase space is sufficiently sampled in the region of interest, the solution can be re-weighted in the later analysis, modeling e.g.~a source with different spectral slope. Also, the continuous injection of pseudo-particles can be constructed by a time integration of the solution (a summation of Green's function, see \cite{Merten-etal-2018} for details), which saves computation time. The latter, however, is only possible when the background fields are stationary.

\subsection{Updated modules}

The \texttt{AdvectionField} modules that provide the background flow $\mathbf{u}$ and its divergence $\nabla \cdot \mathbf{u}$ are extended to take time-dependency into account. The time-dependent advection fields will be publicly available also in the master version of CRPropa 3.2\footnote{See the GitHub repository for details and installation instruction https://github.com/CRPropa/CRPropa3 .}. 
From eq.~(\ref{eq:ODE}) follows that the advection $\vec{u}$ has to be differentiable. This means that the shock needs to be smoothed out with a finite width, e.g.\ with an hyperbolic tangent. As on the other hand, an ideal shock is a discrete jump not only in velocity but also in pressure, density and temperature of the background plasma. This implies constraints on the shock width to obtain the spectra expected at a discrete shock which are discussed in \cite{KruellsAchterberg94, AchterbergSchure2010} and validated for CRPropa in A24 and can be summarized by the following inequality:
\begin{equation}
\label{eq:constraints}
   \left( \left( \nabla \cdot \hat{\kappa} + \mathbf{u} \right) \Delta t \right) \cdot \mathbf{e}_{sh} \leq l_\mathrm{sh} \lesssim \sqrt{2 \hat{\kappa} \Delta t} \cdot \mathbf{e}_{sh} 
\end{equation}
where $\Delta t$ is the simulation time step, $l_{\mathrm{sh}}$ is the shock width. The shock normal is denoted by $\mathbf{e}_{sh}$.

The constraints depend on the simulation time step, to ensure that the resolution is high enough for pseudo-particles to encounter the shock region. Also, from a physical perspective the mean-free path of the particles must be larger than the shock width\footnote{Note, that the shock width in the simulation need not to be equal to the physical width of the shock.}. For the simulation it is sufficient that the diffusive step $\sqrt{2\hat{\kappa} \Delta t}$, which is a measure for that actual stochastic integration step, is larger than the considered shock width. With that, the mean diffusive step for the pseudo-particles can be larger than the mean-free path for a corresponding physical particle, which speeds up the simulation significantly. 

The \texttt{DiffusionTensor} modules are extended to take Bohm diffusion in a magnetic field with varying strength into account. Changing eigenvalues of the diffusion tensor lead to an additional drift-like term in the SDE eq.~(\ref{eq:SDE}) (for details we refer to e.g.~ \cite{KoppEA2012} and MA24).  

We validate transport and acceleration at time-dependent wind profiles by first comparing the spectra at a one-dimensional planar shock in lab frame (where the shock is moving) and shock frame (where the shock speed $v_{\mathrm{sh}} = 0$). We further investigate diffusive shock acceleration (DSA) at a spherical blast wave resulting from a point explosion, using the similarity solution found by Sedov and Taylor as a wind profile \citep{Kahn1975, Sedov-1946, Taylor-1950, Isenberg-1977}. The resulting spectra are compared to an analytical approximation given by \cite{Drury}.

\subsection{1D planar shock}

We show that with the CRPropa simulation using the Euler-Maruyama scheme to integrate the SDE and ODE in eqs.~\ref{eq:SDE} and \ref{eq:ODE} we obtain the same spectra at a one dimensional planar shock moving with $v_{\mathrm{sh}}$ into an undisturbed medium $v_0 = 0$ and at a stationary shock with $u_{\mathrm{sh}} = 0$ and upstream speed $u_0 = - v_{\mathrm{sh}}$. 

\begin{figure}
    \centering
    \includegraphics[width=1\linewidth]{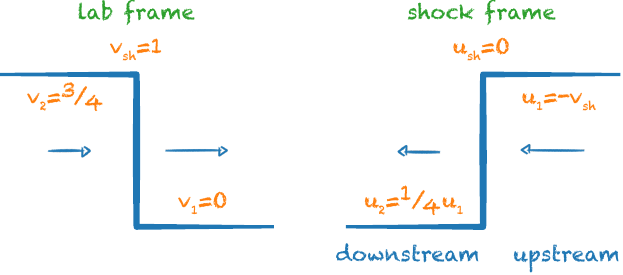}
    \caption{Shock profiles in lab frame (left) and frame in which shock is stationary (right). Assuming a compression ratio $q = 4$ and that the preshock medium does not move ($v_1 = 0$), the postshock speed is $3/4$ of the shock speed. In the shock frame, upstream and downstream side can be specified and the downstream speed equals $1/q$ the upstream speed.  }
    \label{fig:shock-frames}
\end{figure}

The two wind profiles sketched in fig.~\ref{fig:shock-frames} are implemented in the modified version of CRPropa 3.2. In both cases the shock profile is approximated by a $\tanh((x-x_{\mathrm{sh}}/l_{\mathrm{sh}})$ function. The shock width $l_{\mathrm{sh}}$ and time step $\Delta t$ are chosen according to the constraints reported in the previous section. 

The stationary shock is set at $x = 0$, upstream speed\footnote{Computational units e.g.~with $[x] = \mathrm{kpc}$, $[u] = 100\,\mathrm{km/s}$ and $[t] \approx 10\,\mathrm{Myr}$} is $u_0 = 1$ and downstream speed $u_1 = 1/4$. Pseudo-particles are injected at $t = 0$ at $x = 0$ and propagated until the simulation ends at time $T = 100$. Assuming a continuous injection of particles, the solution is obtained by integrating over times $T_i = n \Delta T$ (see \cite{Merten-etal-2018}, A24, MA24 for details). 

In the lab frame, the shock starts at $x = 0$ and moves with constant speed $v_{\mathrm{sh}} = 1$. Pseudo-particles are injected at $t = 0$ in the range $x = [0, 100]$. Since the wind profile is not stationary, the solution cannot be integrated over time, thus, a factor of $n$ more pseudo-particles are injected at $t = 0$. This slows down the simulation by a factor of $n$, however, the uncertainty in the spectra is lower since more independent pseudo-particles are included in the simulation. For better statistics at high energies, we use the \texttt{CandidateSplitting} module (A24, MA24). To compensate for the low number of pseudo-particles at high energies, they are split into two copies once they cross boundaries in energy determined by the expected power-law produced by the shock. In the later analysis, they are weighted according to the number of splits. 

\begin{figure}
    \centering
    \includegraphics[width=1\linewidth]{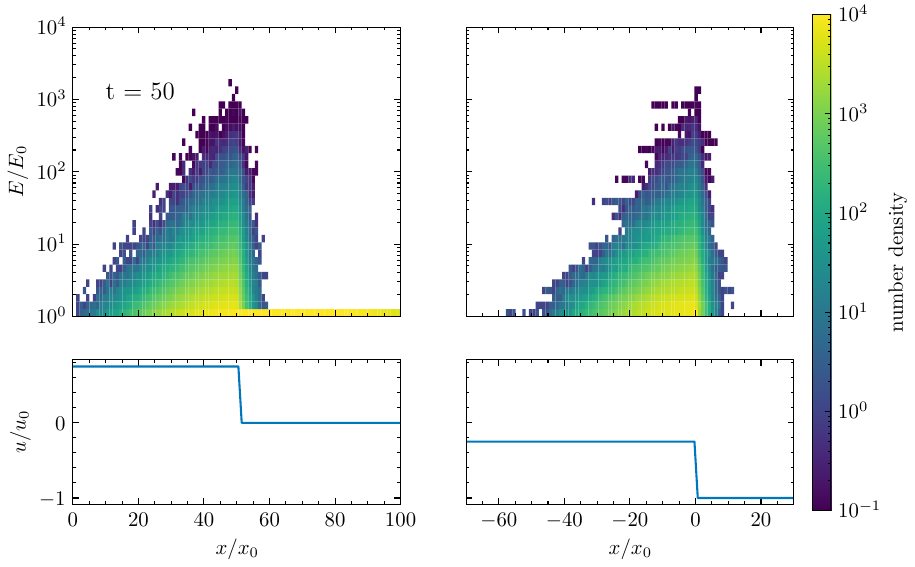}
    \includegraphics[width=1\linewidth]{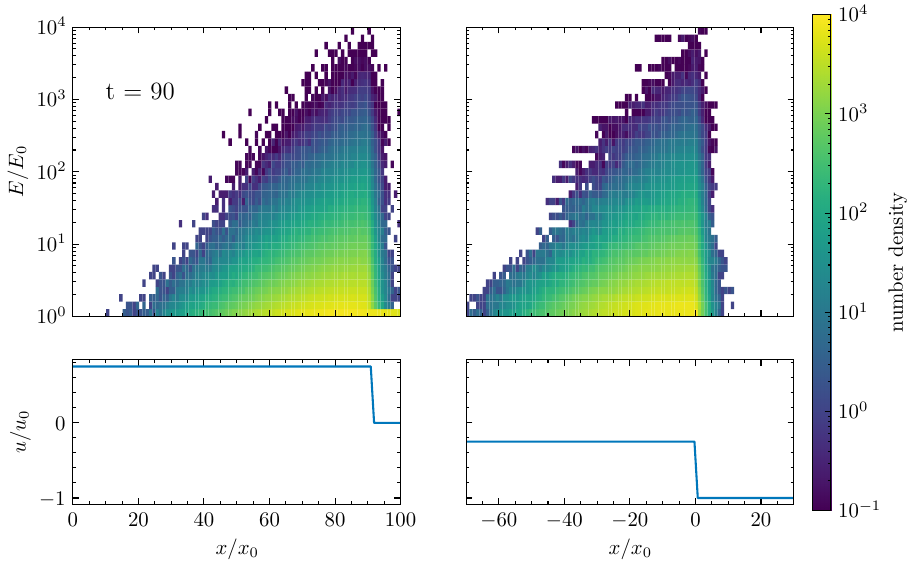}
    \caption{Left: Wind profiles and space-energy histogram in lab frame. Pseudo-particles are injected in the undisturbed medium, the shock is moving through it. Right: Stationary wind profile and space-energy histogram. Pseudo-particles are injected at the shock. For both columns, the top row shows spectra and wind profiles at $T=50$ and the bottom row shows them at $T=90$.}
    \label{fig:Lab-stat-frame}
\end{figure}

Figure \ref{fig:Lab-stat-frame} shows the resulting energy distribution over space at two times $T = 50, 90$. The ultra-relativistic limit $E \approx p\mathrm{c}$ is used. Note, that for the stationary wind the $y$-axis has negative values. The horizontal pattern visible in the histogram for the stationary shock at later times $T = 90$ come from the time integration and have no associated physical effect. For an analysis on the uncertainties coming from the integration in time and the use of the splitting module we refer to MA24. 

\begin{figure}
    \centering
    \includegraphics[width=1\linewidth]{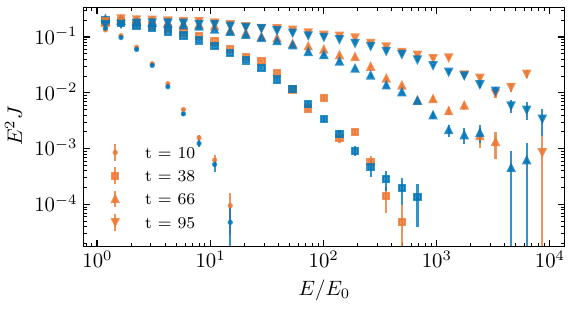}
    \caption{Energy spectra at the shocks in lab frame (moving with the shock) (blue) and in stationary shock frame (orange). The resulting spectral slope and time evolution are the same. Error bars are obtained from the Monte-Carlo error in each bin and proportional to $\sqrt{N}$ where $N$ is the number of independent candidates in the respective bin. The discrepancy mainly is due to the time integration in case of the stationary profile.}
    \label{fig:Lab-stat-spectra}
\end{figure}

Figure \ref{fig:Lab-stat-spectra} shows the energy spectrum $\mathrm{d}\mathcal{N} / \mathrm{d}E \propto E^s$ at the shock, $x = [0,1]$ for the stationary and $x = [x_{\mathrm{sh}}, x_{\mathrm{sh}} + 1]$ moving with the shock in lab frame over time. Both, the expected spectral slope $s = -2$ for a compression of $4$ and the time evolution match very well. To reach the stationary solution over the full energy range one simply needs to run the simulations for longer. 

\subsection{Sedov-Taylor blast wave}
\label{ssec:SedovTaylor}

We extend the model to a spherical explosion. In contrast to the planar case, the postshock medium is now cooled due to expansion and the shock slows down over time. Considering an undisturbed medium with constant density and negligible pressure, a similarity solution provides an analytical wind profile for the blast wave resulting from a point explosion \citep{Sedov-1946, Taylor-1950, Isenberg-1977}. 

The shock radius is given by
\begin{equation}
    R(t) = \left(\frac{E_{\mathrm{inj}}}{\rho_0}\right)^{1/5} t^{2/5}\quad,
\end{equation}
and only depends on the energy in the explosion $E_{\mathrm{inj}}$ and undisturbed density $\rho_0$. From the shock radius follows the shock speed
\begin{equation}
     \dot{R} = \frac{2}{5}\left(\frac{E_{\mathrm{inj}}}{\rho_0}\right)^{1/5} t^{-3/5}\quad.
\end{equation}
With that the wind profile is given by
\begin{equation}
\label{eq:STwind}
    u(r) = \left \lbrace \begin{array}{lcl}
        V(\xi)\xi \dot{R} &, \quad & 0 < \xi < 1  \\
         0 &, \quad & \xi > 1
    \end{array} \right.
\end{equation}
where $\xi = r / R(t)$ and $V(\xi) = 3(\xi^8 + 1)/(3 \xi^8 + 5)$ \citep{Kahn1975, Drury}.\footnote{Note the different notation used in the works by Kahn and Drury. The expression in eq.~(7.23) in \citet{Kahn1975} is equivalent to eq.~(\ref{eq:STwind}) given here and in \citet{Drury}.} The upstream speed equals the shock speed, $u_1 = \dot{R}$, and $u_2 = \dot{R} (1 - V(1))$ is the downstream speed. Note that $V(1) = 3/4$, analogous to the postshock speed in the planar case for a compression ratio of $4$. 

\begin{figure}
    \centering
     \includegraphics[width=1\linewidth]{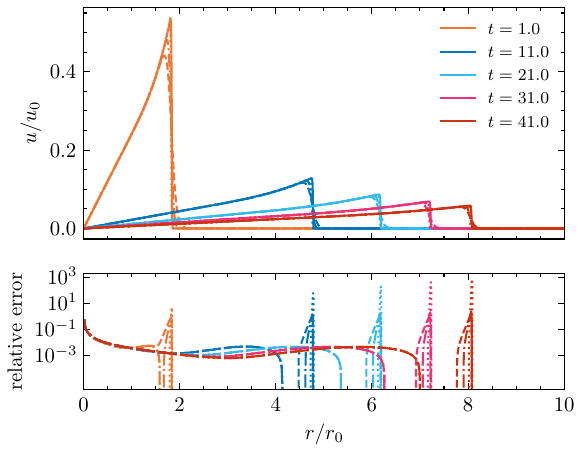}
    \caption{Top: Wind profile of the self-similar Sedov-Taylor solution. The shock slows down over time $\propto t^{-3/5}$. Here $E_{\mathrm{inj}} = 20$, $\rho_0 = 1$. Athena++ simulation (solid line) is compared to analytical profile with different shock widths, $l_{\mathrm{sh}} = [0.1, 0.05, 0.01]$ (dashed, dash-dotted, dotted line). Bottom: Relative error between the Athena++ simulation and approximated analytical profile. For the smallest shock width the error increases over time since the shock position of the analytical solution and Athena++ simulation diverge. }
    \label{fig:STwindprofile}
\end{figure}

Equation (\ref{eq:STwind}) is not differentiable at the shock, $\xi = 1$. Just like in the previous section a smooth wind profile is modeled by an hyperbolic tangent at the shock radius.  We validate this approach by comparing the modified Sedov-Taylor solution with a velocity profile resulting from a simulation with the hydrodynamic code Athena++ \citep{Athena++}. The initial values are chosen to be in agreement with the assumption for the analytical solution: $u_0 = 0$, $p_0 \ll 1$, $\rho_0 = 1 = \mathrm{const}$. The energy $E_{\mathrm{inj}}$ is injected in a small volume around the origin $r_0 \ll 1$ and conserved during the simulation. In fig.~\ref{fig:STwindprofile} the resulting wind profile and modified analytical profiles are shown over time. These simulations are one-dimensional, only allowing radial motion. However, the polar and azimuthal bounds cover the full $4\pi$ solid angle. Including the full range of angles is the only way to get profiles which agree with the classic and modified Sedov-Taylor solution. 
The relative error between the approximation and simulation is shown in the bottom panel of fig.~\ref{fig:STwindprofile}. The profiles agree well, except right at the shock which is expected due to the smoothing of the hyperbolic tangent. However, especially for the smallest shock width, the error increases over time. Here, the shock positions of the approximation and simulation do not exactly agree which causes a strong deviation right at the shock. This does not impact the acceleration, as long as the compression ratio at the shock remains the same and the shock width together with the time step fulfills the constraints discussed previously. 
 
When simulating particle acceleration at such a blast wave, the inequality \ref{eq:constraints} has to be fulfilled in the frame in which the shock is stationary ($\dot{R} = 0$). For efficient acceleration, particles have to be confined and their diffusion coefficient needs to be small compared to the shock radius $R$ and speed $\dot{R}$. 

We first consider the case of a constant diffusion coefficient $\kappa$. 
The maximal energy that can be reached at a given time can be calculated from \citep{LagageCesarsky83, Drury}
\begin{equation}
\label{eq:Lagage}
    \frac{\mathrm{d}p}{\mathrm{d}t} = \frac{u_1 - u_2}{3} \frac{p}{\kappa_1 / u_1 + \kappa_2 / u_2}, 
\end{equation}
where $u_1 = \dot{R}$, $u_2 = \dot{R}/4$ and, here, $\kappa_1 = \kappa_2 = \kappa$. With the rather unphysical condition of a constant diffusion coefficient, the maximal energy grows fast and particle feedback to the plasma cannot be neglected. The simulation is instead validated by comparing the estimated spectral slope. In the following section, the maximal energy is obtained considering Bohm diffusion.

When the shock radius $R$ is large compared to the diffusion length scale $\kappa / u$ the shock can be seen as a planar shock and we expect the energy spectrum at the shock to have the characteristic slope $s = -2$. However, with increasing diffusion coefficient, the curvature of the shock has to be taken into account. In the limit where  $\epsilon = \kappa / (\dot{R} R)$ is still a small parameter \cite{Drury} gives an approximation of the spectral slope $a$ at the shock
\begin{equation}
\label{eq:slope}
    a = 4 \left[ 1 + (3 + b) \left( \frac{\kappa_1}{R u_1} + \frac{\kappa_2}{R u_2} \right) + \frac{\kappa_2}{R u_2} \right] + O(\epsilon^2) \quad,
\end{equation}
where $b$ describes the injection of particles, assuming $f \propto p^{-a} R(t)^b$. Note, that for $\epsilon \rightarrow 0$, the approximation yields the slope at a planar shock, $f \propto p^{-4}$. The small deviations from the planar solution originate from the finite rate at which particles are injected and accelerated (the acceleration time scale is proportional to $(\kappa_1/(R u_1) + \kappa_2/(R u_2))$) as well as the adiabatic cooling in the expanding wind behind the shock (determined by the downstream diffusive length scale $\kappa_2 / u_2$).

At $t = 0$ the shock speed diverges and with that the wind profile given by eq.~(\ref{eq:STwind}) at $r = 0$. However, the blast wave approximation is only meant to be applied at late times, once the ejecta have swept out there own mass and the explosion can be treated as coming from a point source. At early times, the shock speed is high and acceleration is fast, as long as particles are confined by the shock. When simulating with CRPropa 3.2, acceleration is efficient when the inequality (\ref{eq:constraints}) involving the time step, shock width and diffusive step are fulfilled. Therefore the time step is chosen such that this is only true for $t > 1$, when the shock speed decreases to a reasonable speed.

\begin{figure}
    \centering
    \includegraphics[width=1\linewidth]{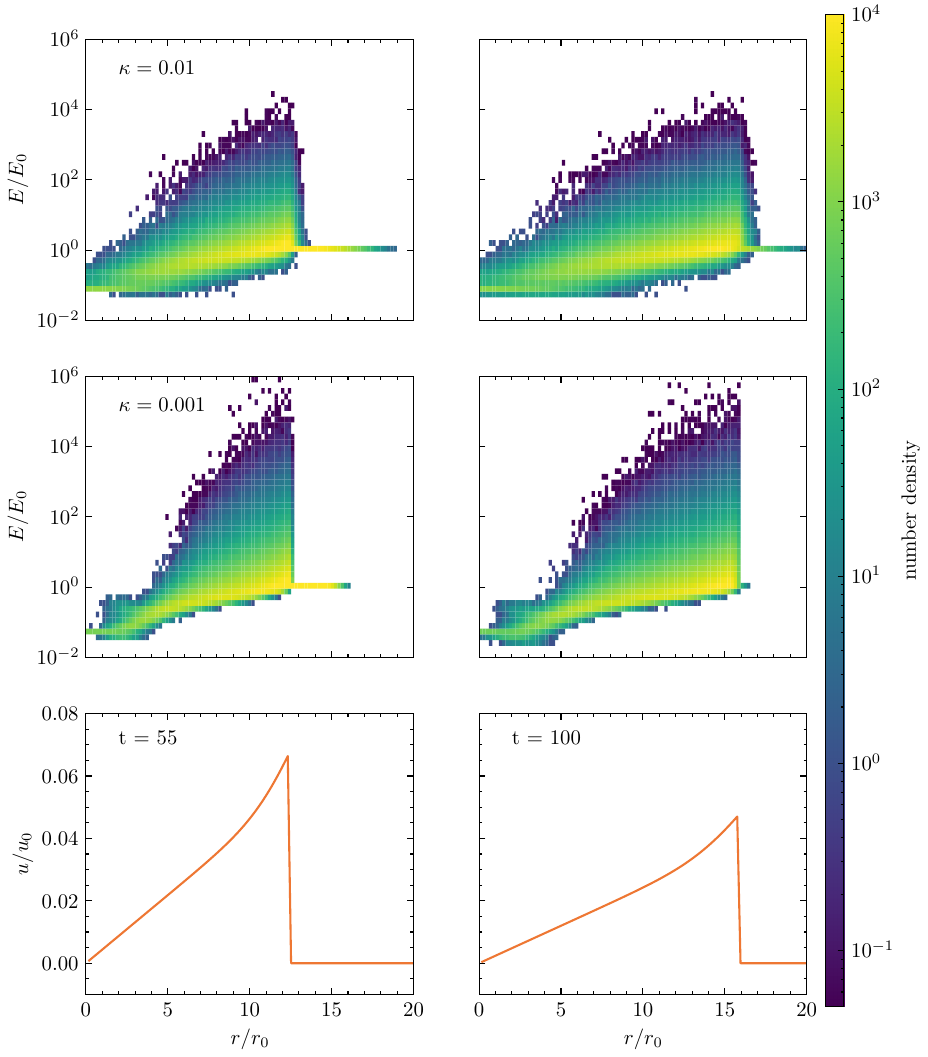}
    \caption{Acceleration at Sedov-Taylor blast wave. Particles are accelerated at the shock and are cooled in the expanding wind.}
    \label{fig:Sedov_acc}
\end{figure}

Figure \ref{fig:Sedov_acc} shows the resulting distribution $fp^2$ at $t = 55$ and $t = 100$, with the simulation parameters: $E_{\mathrm{inj}} = 100$, $\rho_0 = 1$, $\kappa = 0.01$ and $\kappa = 0.001$, $l_\mathrm{sh} = 0.01$. A number of $10^6$ pseudo-particles with $p = p_0$ is injected homogeneously in the undisturbed medium at $t = 0$. Particles gain energy at the shock and can lose energy in the expanding wind behind the shock. Acceleration is fast at early times, when the shock speed is high, it slows down over time, so that the maximal 
energies reached at $t = 55$ and $t = 100$ are similar. This is in agreement with eq.~(\ref{eq:Lagage}). 

\begin{figure}
    \centering
    \includegraphics[width=1\linewidth]{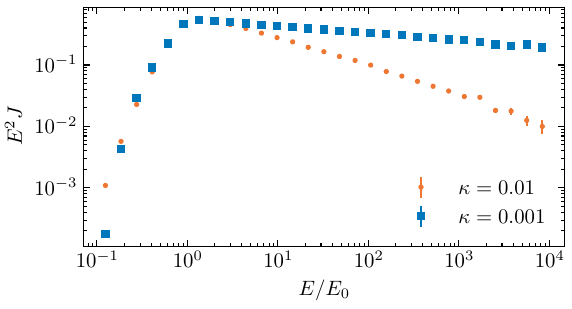}
    \caption{Spectra at the shock at $T = 100$ for $\kappa = 0.01, 0.001$. The spectral slope matches with the approximation given by Drury: $s(\kappa = 0.001) \approx 2.075$, $s(\kappa = 0.01) \approx 2.75$. }
    \label{fig:Sedov_spectrum}
\end{figure}

Figure \ref{fig:Sedov_spectrum} shows the energy spectra at the shock at $t = 100$.  With sufficiently small diffusion coefficient eq.~(\ref{eq:slope}) yields $a \approx 4$, the spectrum at a planar shock. With higher diffusion coefficient, the resulting spectrum is steeper which is in agreement with eq.~(\ref{eq:slope}). For energy-dependent diffusion, this effect leads to a steepening of the spectral slope with increasing energy. The spectrum is affected by the finite size and acceleration time of the shock --- a Hillas effect --- and the adiabatic cooling in the downstream region of the shock due to the expanding wind. In case of a spherical termination shock there is a similar effect: Particles are cooled upstream (which is inside) which leads to an enhancement in the energy spectrum at high energies right before the cut-off \citep{Drury, FlorinskiJokipii2003}.  

\section{Individual spherical shocks: Acceleration and escape back to the Galaxy}
\label{sec:sphericalwind}

Strong supernova explosions or superbubbles could lead to shocks propagating out of the Galaxy \citep{Tomisaka-Ikeuchi-1986, Dorfi-etal-2012} and it has been proposed proposed that CRs could be accelerated at those local structures \citep{Dorfi-etal-2019}. Like a potential Galactic wind termination shock, such outbursts would contribute to the CR spectrum in the transition region between the spectral knee and ankle \cite{JokipiiMorfill87, BustardEtAl2017, Merten-etal-2018, MukhopadhyayEtAl2023}. Although a Galactic termination shock is larger and could confine CRs at higher energies, the time-dependent structures are closer to the Galactic disk, making escape back to the Galaxy more promising. While \cite{Dorfi-etal-2012} and \cite{Dorfi-etal-2019} give an estimate on the maximal energy that could be reached at their shock structures, they do not discuss the probability of CRs returning back and how the transport affects the shape of the spectrum. For the termination shock, the escape back to the Galaxy would be upstream, in this scenario, downstream. However, the downstream frame is now moving with the (time-dependent) shock speed out of the Galaxy. All this combined suggests a non negligible influence on the accelerated spectrum $s=-2$.

We only expect shocks strong enough for particle acceleration when they run out of the Galactic disk and the density decreases. We refer to \citet{Weaver-etal-1977} for an estimate of the shock speed considering the constant density of the interstellar medium. Thus, starting from the Sedov-Taylor blast wave in sec.~\ref{ssec:SedovTaylor}, the wind profile is modified to take a decreasing density profile $\rho \propto 1/r^2$ into account. Furthermore, we consider a more realistic, energy-dependent diffusion coefficient compared to sec.~\ref{ssec:SedovTaylor}. When the mean free path increases with energy, re-accelerated high energy CRs have a higher probability to propagate back to the Galaxy. On the other hand, DSA is most effective for small diffusion coefficients. Thus, the Bohm limit is used in the following to evalute the most promising scenario for a contribution to the flux in the knee to ankle region. 

\subsection{Wind profile: Sedov-Taylor with decreasing density}

\begin{figure}
    \centering
    \includegraphics[width=1\linewidth]{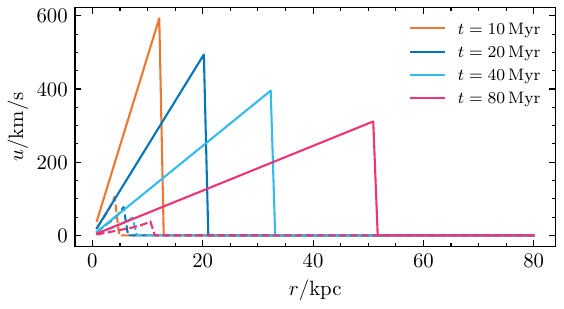}
    \caption{Analytical solution of the Sedov-Taylor blast wave with constant density (dashed line) compared to the Isenberg solution with decreasing density profile (solid line) for $E_\mathrm{inj} = 10\,E_{\mathrm{SN}}$, $r_0 = 1\,\mathrm{kpc}$ and $\rho_0 = 1 \,\mathrm{m_p} / \mathrm{m}^3$. Considerably higher shock speeds are reached for the latter. 
    }
    \label{fig:ST_comparison}
\end{figure}

Outside the Galactic disk the density decreases and the shock can survive longer and propagate further out with higher velocities compared to the constant density scenario. While the original Sedov-Taylor solution assumes a constant density in the undisturbed medium, \cite{Isenberg-1977} found a solution for $\rho_0(r) = A r^{-2}$. The velocity profile then reads
\begin{equation}
    u(r) = v_s V_0(\lambda), r < R
\end{equation}
with $v_s = \mathrm{d}R/\mathrm{d}t = 2 / 3 (E_\mathrm{inj} / A)^{1/3} t^{-1/3}$ and 
\begin{equation}
    V_0 = 2 / (\gamma + 1) \lambda,
\end{equation}
where $\gamma = 5/3$ and $\lambda$ is the dimensionless parameter 
\begin{equation}
    \lambda = \left( \frac{A}{E_\mathrm{inj} t^2} \right)^{1/3} r,
\end{equation}
of the similarity solution. Figure \ref{fig:ST_comparison} shows the resulting profile with $A = \rho_0 r_0^2$ with $r_0 = 1\,\mathrm{kpc}$ and $\rho_0 = 1 \,\mathrm{m_p} / \mathrm{m}^3$, where $\mathrm{m_p}$ is the proton rest mass, in comparison to the one with constant density $\rho_0$. The density is significantly higher for $r < r_0$, thus, the value of $\rho_0$ roughly leads to a reasonable density (specifically, this value of $\rho_0$ gives $\rho_0(r=1\,\mathrm{pc}) = 1 \mathrm{m_p}/ \mathrm{cm}^3$). The injected energy is $E_\mathrm{inj} = 10\,E_{\mathrm{SN}} = 10^{52}\,\mathrm{erg}$. For comparison, we also plot the Sedov-Taylor solution for constant density in fig.~\ref{fig:ST_comparison}. In addition to making it to larger radii, the shocks have larger speeds. 

While such a wind profile is realistic for large radii, the density is overestimated for small $r$ and diverges for $r \rightarrow 0$. Thus, we expect it only to be valid at the larger scales, which are the relevant ones for the study here. Like in sec.~\ref{ssec:SedovTaylor} the constraints on the time step and shock width are chosen such that it is fulfilled for $t > 1\,\mathrm{Myr}$.

\subsection{Diffusion: Bohm diffusion at parallel shock}

DSA is most efficient for small diffusion coefficients. We consider strong turbulence close to the shock and Bohm diffusion coefficient following $\kappa(p,r) = p \mathrm{c}^2 / (3 q B(r))$. For efficient acceleration we assume the spherical shock to be parallel, thus, the magnetic field is radial at the shock. The magnetic field strength $B$ decreases with $1/r^2$, hence, the Bohm diffusion coefficient increases. We set the magnetic field strength to $B_0 = 10\,\mu\mathrm{G}$ at $r_0 = 1\,\mathrm{kpc}$.

The spatial dependence of the Bohm diffusion coefficient leads to a drift-like term in the SDE:
\begin{equation}
    \frac{\partial \kappa}{\partial r} = \frac{2 p \mathrm{c}}{3 q B(r)} \frac{1}{r}
\end{equation}
which is implemented in the \texttt{DiffusionTensor} module and must be taken into account when constraining the time step (eq.~(\ref{eq:constraints})). Depending on the energy and magnetic field strength, the drift-like term can get larger than the advection\footnote{It is, however, questionable, if the diffusive regime is then still applicable.}.  

\begin{figure}
    \centering
    \includegraphics[width=1\linewidth]{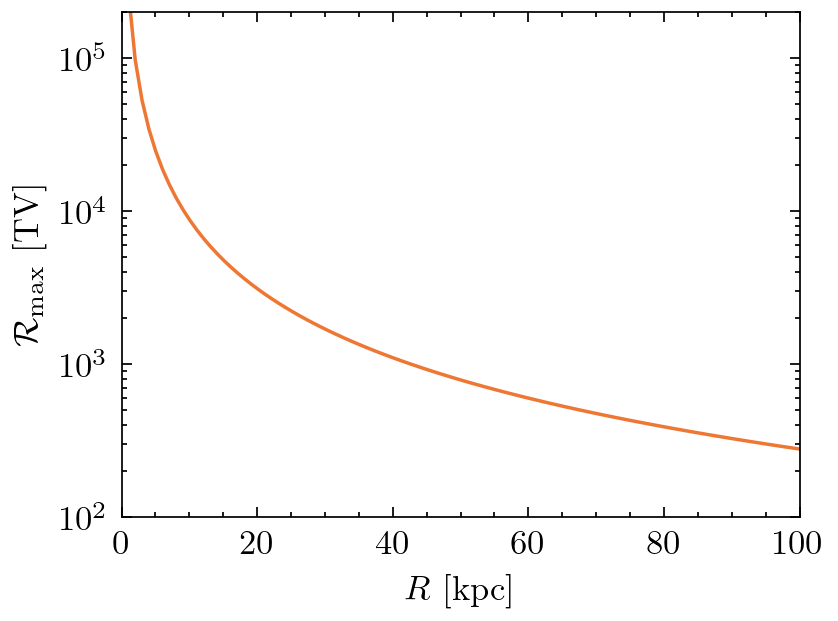}
    \caption{To experience DSA, the diffusive length scale $\kappa / u $ must be small compared to the shock radius $R$. With $u = \dot{R}$, the maximal diffusion is given by $\kappa(p,r) = R \dot{R}$. With the shock slowing down and the diffusion coefficient increasing with $r$, the rigidity that is still confined at the scales of the shock decreases.}
    \label{fig:maxrig}
\end{figure}

We expect particles to be diffusive on the scales of the shock, when the diffusive length scale is smaller than the shock radius, $\kappa / \dot{R} < R$. With $E_\mathrm{inj} = 10\,E_{\mathrm{SN}}$ and the previously described density profile and magnetic field strength, the maximal rigidity $\mathcal{R} = E/q$ that is still confined can be calculated and is shown in fig.\ref{fig:maxrig} depending on the shock radius $R$. Similar considerations have been made by e.g.~\citet{LagageCesarsky83} or \citet{Drury}.

\subsection{Estimate of the maximal energy}

With the shock speed and diffusion coefficient the maximal energy that can be reached at a time $t$ is estimated by integrating eq.~(\ref{eq:Lagage})
\begin{equation}
    \label{eq:max_energy}
    \frac{\mathrm{d}p}{\mathrm{d}t} = \frac{3 q B(R)}{20} \dot{R}^2,
\end{equation}
 where the magnetic field is evaluated at the shock radius and $R$ and $\dot{R}$ depend on time. With $C = q  \mathrm{c} B_0 r_0^2 / 15$ and $E \approx p\mathrm{c}$ follows
\begin{equation}
    E_\mathrm{max}(t) = C \left[ \frac{1}{t_0} - \frac{1}{t}\right] + E_0,
\end{equation}
which is independent on the injected energy $E_{\mathrm{inj}}$ and with that the shock speed. This is due to the decreasing magnetic field: With higher shock speed, radii with low magnetic field are reached faster and the acceleration is slowed down. Thus, for the considered wind profile the effect of a higher shock speed on the acceleration time is canceled out by the higher diffusion. 

\begin{figure}
    \centering
    \includegraphics[width=1\linewidth]{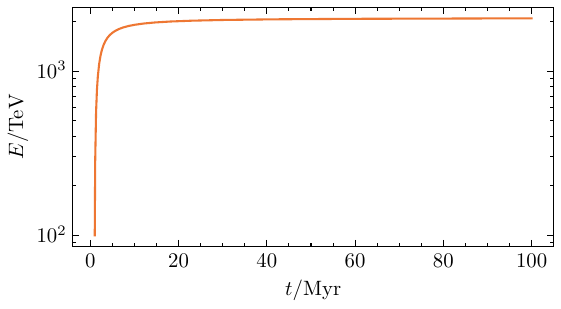}
    \caption{Estimate of the maximal energy that protons reach at the shock over time for $B_0 = 10\,\mu\mathrm{G}$, $r_0 = 1\,\mathrm{kpc}$, $t_0 = 1\,\mathrm{Myr}$ and $E_0 = 100 \,\mathrm{TeV}$.}
    \label{fig:max_energy}
\end{figure}

The resulting maximal energy for $B_0 = 10\,\mu\mathrm{G}$, $r_0 = 1\,\mathrm{kpc}$, $t_0 = 1\,\mathrm{Myr}$ and $E_0 = 100 \,\mathrm{TeV}$ is shown in fig.~\ref{fig:max_energy} and goes up to $2\,\mathrm{PeV}$. 

\subsection{Simulation set-up}

\texttt{Candidates} are injected at $t = 0$ with an energy spectrum $E^{-1}$ from TeV to $100\,$TeV in the volume $r < 60\,\mathrm{kpc}$. From fig.~\ref{fig:maxrig} it is clear, that for $r >  60\,\mathrm{kpc}$ re-accelerated particles with higher energy than PeV are not diffusive on the scales of the shock and are not well described in the ensemble-averaged regime. The injected energy spectrum is re-weighted in the later analysis to $E^{-2}$. With $E_{\mathrm{inj}} = 10\,E_\mathrm{SN}$, $r_0 = 1\,\mathrm{kpc}$ and $\rho_0 = 1 \,\mathrm{m_p} / \mathrm{m}^3$ the expanding shock reaches a radius of $ 60\,\mathrm{kpc}$ at $t \approx 80\,\mathrm{Myr}$. All \texttt{Candidates} are protons, since interactions are not considered the result can be applied to heavier particles with the same rigidity as well.
The \texttt{CandidateSplitting} module is used to increase statistics at high energy. 

To investigate the effect of the shock speed and the magnetic field strength on the maximal energy and transport back to the Galaxy, three simulations are performed: [$E_{\mathrm{inj}} = 10\,E_\mathrm{SN}$, $B_0 = 10\,\mu\mathrm{G}$], [$E_{\mathrm{inj}} = 100\,E_\mathrm{SN}$, $B_0 = 10\,\mu\mathrm{G}$], [$E_{\mathrm{inj}} = 10\,E_\mathrm{SN}$, $B_0 = 100\,\mu\mathrm{G}$]. 

\subsection{Adaptive step: Energy-dependent shock width}

The small Bohm diffusion coefficient, $\kappa(\mathrm{TeV}, r_0) =1.5\times 10^{23}\mathrm{cm}^2 \mathrm{s}^{-1}$, implies a small shock width $l_{\mathrm{sh}}$, so that the mean diffusive step is larger than the shock width. On the other hand, a small shock width affords a small time step, since the advective step has to be smaller than the shock width. At $t = 1\,\mathrm{Myr}$ and $r = R(1\,\mathrm{Myr})$, a time step of $\Delta t = 10^{-5}\,\mathrm{Myr}$ fulfills the constraints considering pseudo-particle energies of TeV. We expect re-acceleration up to $10\,\mathrm{PeV}$, thus, the diffusive step increases by a factor larger than $1000$. However, the diffusive step has to be on the order of the shock width, otherwise pseudo-particles do not encounter the shock (\citet{KruellsAchterberg94}, A24). 

The previously implemented adaptive step (A24, MA24), scales the time step down to keep the diffusive step within bound. Here, the time step is already quite small, decreasing it further would lead to high computation times. Instead, we scale the shock width with the square root of the pseudo-particles energy. At this point we emphasize again, that the shock width does not have to be equal to the physical width of the shock. 

\subsection{Simulation results}

\begin{figure*}
    \resizebox{\hsize}{!}
            {\includegraphics[width = 2\textwidth]
    {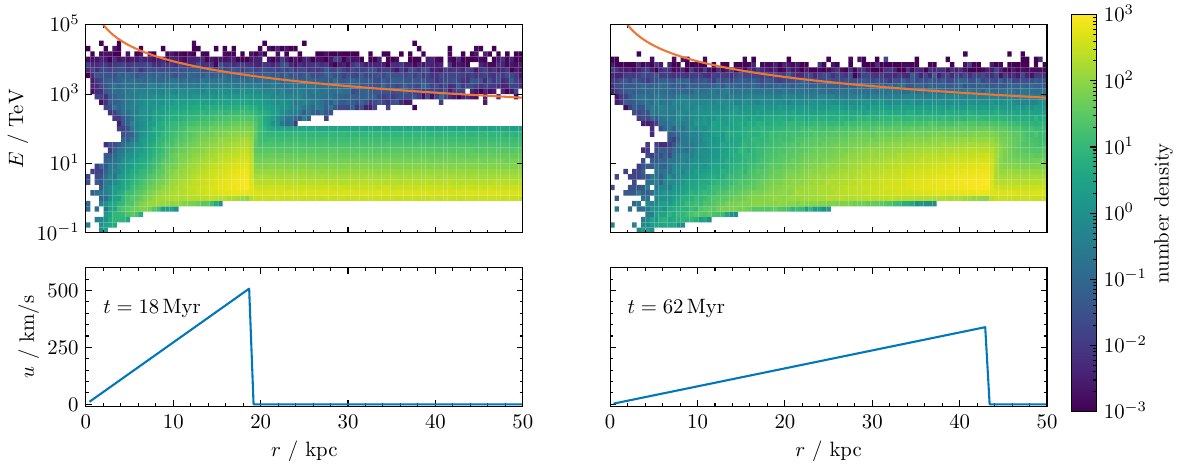}}
    \caption{Wind profiles and space-energy histogram at $t = 18\,\mathrm{Myr}$ and $t = 62\,\mathrm{Myr}$ for $E_0 = 10\,E_\mathrm{SN}$, $B_0 = 10\,\mu\mathrm{G}$. Particles are accelerated close to the Galaxy, when the shock speed is high and diffusion low. The orange line indicates the maximal energy protons are still diffusive. }
    \label{fig:Sim1_hist}
\end{figure*}

\begin{figure}
    \centering
    \includegraphics[width=1\linewidth]{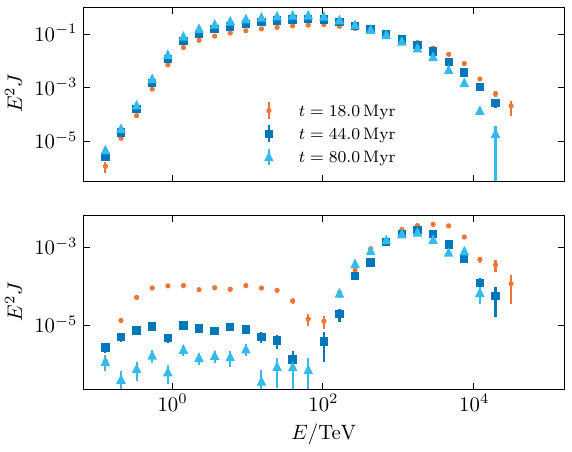}
    \caption{Energy spectra over time at the moving shock (top) and at $5\,\mathrm{kpc}$ (bottom) for $E_0 = 10\,E_\mathrm{SN}$, $B_0 = 10\,\mu\mathrm{G}$. Only particles with sufficiently high energy make it back to the Galaxy.}
    \label{fig:Sim1_spectra}
\end{figure}

Figure \ref{fig:Sim1_hist} shows the resulting number density $fp^2$ at two times. Particles are accelerated up to $50\,\mathrm{PeV}$. High energy particles experience high diffusion and drift $\partial \kappa / \partial r$ and quickly move out of the Galaxy. Particles above the orange line do no longer fulfill the criterion $\kappa < R\dot{R}$, meaning they are not properly described by the diffusion \textit{ansatz}. Since in the further analysis (see fig.~\ref{fig:Sim1_spectra}) we focus on the particle spectra at the shock we do not expect a large difference on short time scales between the diffusive particle population and the ballistic one. The other observer position is at $r=5\,\mathrm{kpc}$, where almost all particles are diffusive for all times.

Figure \ref{fig:Sim1_spectra} shows the spectra at the moving shock at different times. It can be divided in three parts: The injected energy range $E_{\mathrm{inj}} = [\mathrm{TeV}, 100\,\mathrm{TeV}]$, the cooled energy range $E_{\mathrm{cool}} < \mathrm{TeV}$, and the re-accelerated range $E_{\mathrm{acc}} > 100\,\mathrm{TeV}$. The injected energy range flattens over time, which is an artifact of the finite injection range: low energy particles are accelerated up to higher energies and increase the flux. Without the minimum injected energy, the loss of accelerated particles at such energies would be compensated by the acceleration of particles from even lower energies. Cooling also flattens the spectrum: high energy particles on average spend less time in the expanding wind. This is well-visible in the cooled energy range. 

The re-accelerated energy range has a spectral slope $ < -2$. For energy-dependent diffusion follows from eq.~(\ref{eq:slope}) that the spectrum steepens with increasing energy. Note, that this occurs even below the cut-off energy due to finite acceleration time and size of the accelerator. For energy $> 100\,\mathrm{TeV}$ the steepening of the spectrum is well visible in fig.~\ref{fig:Sim1_spectra}.

Interestingly, the energy spectrum at the shock also slightly steepens over time. The shock slows down and, thus, accelerates particles to a steeper spectrum. Particles are cooled in the downstream region of the shock. When the shock is not fast enough to compensate for the energy losses anymore, the spectrum steepens. Also, the shock propagates in an area with lower magnetic field and without further magnetic field amplification, the diffusion coefficient increases. Taking both effects into account, eq.~(\ref{eq:slope}) predicts a steepening of the spectrum with time. 

The bottom panel of fig.~\ref{fig:Sim1_spectra} shows the energy spectrum at $r = 5\,\mathrm{kpc}$ for an estimate of the particle spectrum that escapes back to the Galaxy. It can be divided into the injected part that is cooled and advected outwards over time, and the re-accelerated part that is able to diffuse back. While the low energy part decays over time, the re-accelerated part is less affected by advection and is almost stationary. 

\begin{figure*}
    \resizebox{\hsize}{!}
            {\includegraphics[width = 2\textwidth]{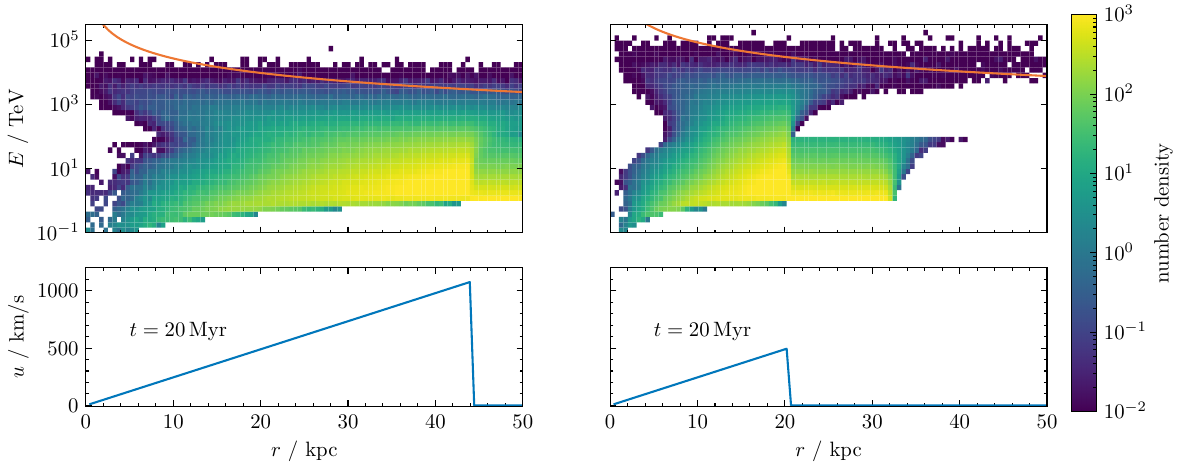}}
    \caption{Wind profiles and space-energy histogram at $t = 20\,\mathrm{Myr}$ for $E_0 = 100\,E_\mathrm{SN}$, $B_0 = 10\,\mu\mathrm{G}$ (left) and for $E_0 = 10\,E_\mathrm{SN}$, $B_0 = 100\,\mu\mathrm{G}$. The orange line indicates the maximal energy protons are still diffusive for the respective magnetic field and shock speed.}
    \label{fig:Sim23_hist}
\end{figure*}

\begin{figure*}
   \resizebox{\hsize}{!}
            {\includegraphics[width = 2\textwidth]{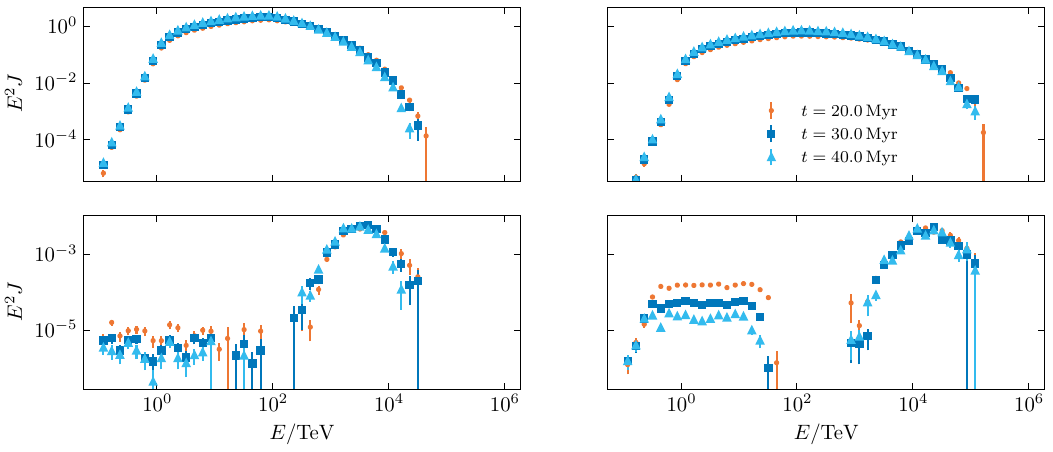}}
    \caption{Energy spectra over time at the moving shocks (top) and downstream at $5\,\mathrm{kpc}$ at for $E_0 = 100\,E_\mathrm{SN}$, $B_0 = 10\,\mu\mathrm{G}$ (left) and $E_0 = 10\,E_\mathrm{SN}$, $B_0 = 100\,\mu\mathrm{G}$ (right).}
    \label{fig:Sim23_spectra}
\end{figure*}

We further investigate the impact of the shock speed by increasing the energy of the explosion to $E_0 = 100\,E_\mathrm{SN}$ (left column of fig. \ref{fig:Sim23_hist} and \ref{fig:Sim23_spectra}) and the impact of the diffusion coefficient by amplifying the magnetic field to $B_0 = 100\,\mu\mathrm{G}$ (right column of fig. \ref{fig:Sim23_hist} and \ref{fig:Sim23_spectra}). The magnetic field can be amplified due to the Bell instability driven by the CRs themselves \citep{Bell-2004, Amato-Blasi-2009}. A high magnetic field is key to confine and accelerate CRs at shocks and it has been proposed that the Bell instability leads to significant amplification \citep{Zweibel-Everett-2010, Drury-Downes-2012}. 

With a higher injected energy the wind speed increases. Like predicted in eq.~(\ref{eq:max_energy}) the maximal energy reached at the shock is the same $\approx 10\,\mathrm{PeV}$. Less low energy particles make it back to the Galactic boundary since they are advected out with the faster flow. In this scenario, the maximal energy is independent on the individual energy in the outbursts. If, however, the magnetic field strength or the density decrease with a different dependency on $r$, the maximal energy would change. 

With a smaller diffusion coefficient, particles are accelerated to a harder spectrum and reach higher maximal energy. However, because of the decreased diffusion coefficient, the particles are advected outwards more efficiently. Therefore, fewer particles make it back to the Galaxy. 

\section{Consecutive planar shocks: Softer spectra and collisions}
\label{sec:twoshocks}

With explosions like the one discussed in the previous section happening frequently in the Galaxy, a stationary flux could develop. Also, shocks are likely to run into each other and their interaction would result in complex landscapes affecting CR acceleration and their transport back to the Galaxy. 

We approach the idea of CR acceleration at interacting shocks by considering two one-dimensional planar shocks, one emerging after the other. From the Rankine-Hugoniot conditions follows, that the second shock has to be is faster than the first shock: The shock has to be faster than the speed of sound $c_{\mathrm{s}}$ in the preshock medium, otherwise it does not fulfill the Mach criterion $M > 1$, with $M^2 = u^2/c_{\mathrm{s}}^2$. When the preshock medium is already shocked and in motion, it follows 
\begin{equation}
    \label{eq:RHvs2}
    \left(v_{\mathrm{sh}, 2} - 3/4\,v_{\mathrm{sh}, 1}\right)^2 >  c_{\mathrm{s}, 1}^2 \quad,
\end{equation}
where 
\begin{equation}
c_{\mathrm{s}, 1}^2 = \gamma \frac{p_1}{\rho_1} = \gamma \frac{3}{4} \frac{\rho_0 u_0^2}{\rho_1} = \frac{5}{16} v_{\mathrm{sh}, 1}^{2} \quad,
\end{equation}
assuming a non-relativistic fluid with $\gamma = 5/3$. For the speed of the second follows from eq.~(\ref{eq:RHvs2}): $v_{\mathrm{sh}, 2}   \geq(3+\sqrt{5})/4 \sim  1.309\,v_{\mathrm{sh},1}$. Note, that also the compression of the shock changes, likely being lower than $4$. 

For secondary supernova or suberbubbles explosions, this means that their energy must be significantly higher to produce a flow which is a second shock. When radiative cooling or spherical expansion (e.g. a Sedov-Taylor blast wave) is taken into account this constraint becomes less severe, since the velocity/density behind the first shock can be lower. \cite{Dorfi-etal-2019} confirmed with their hydrodynamic simulation, considering a flux tube geometry, that secondary shocks are always faster than the previous shocks and eventually collide.

\subsection{Wind profile: Colliding shocks}

\begin{figure}
    \centering
    \includegraphics[width=\linewidth]{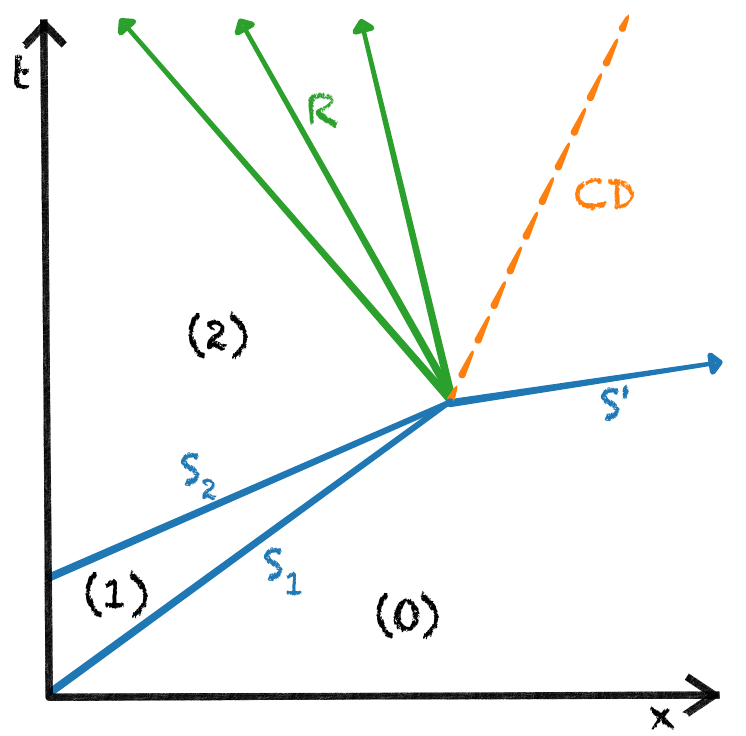}
    \caption{Sketch of the colliding shock problem: Two shocks start at $x = 0$ at different times, the second shock is faster than the first and catches up. When they collide a new Riemann problem forms determined by the states ($0$) and ($2$). Three new regions develop: The shocked medium between the single shock and the contact discontinuity (CD), the medium between the CD and the rarefaction wave (R), and the rarefaction wave itself. Pressure and velocity do not change over the CD. The rarefaction wave travels 
    at the sound speed.}
    \label{fig:ShockCollision}
\end{figure}

We consider two one-dimensional planar shocks that are propagating in the same direction. What happens when they collide? A new Riemann problem emerges, with the undisturbed medium ($0$) on one side and the twice shocked medium ($2$) on the other. From \cite{CourantFriedrichs} we know that a contact discontinuity forms and a rarefaction wave travels backwards as sketched in fig.~\ref{fig:ShockCollision}. 


For more quantitative statements, e.g.~on the speed of the new shock, the scenario can be viewed as a modification of the Sod shock tube problem \citep{1978Sod}. It is a widely used test case for hydrodynamic codes, where the initial condition for density and pressure on one side of a one-dimensional domain are chosen high but low on the other side. The pressure and density gradient drive a shock wave when the simulation starts. The original problem from \cite{1978Sod} has set values for each quantity and region. For example, the speed on both sides is set to zero. In the Riemann problem we aim to solve, the high pressure/high density medium is already in motion. Therefore, we build a two-shock initial condition from the generalized shock tube problem already in Athena++.

The (single) shock tube problem can be solved analytically with the Rankine-Hugoniot conditions; however, the pressure and velocity on both sides of the contact discontinuity have to be calculated iteratively. Instead, we set up the modified shock tube problem in Athena++ to obtain pressure, density and velocity in the unknown region\footnote{Note, that the density changes over the contact discontinuity, but not the pressure/velocity.}. Those simulated values are used to construct the analytical profile for the full velocity profile and its derivative, which is necessary for modeling particle acceleration with CRPropa. 

\begin{figure}
    \centering
    \includegraphics[width=1\linewidth]{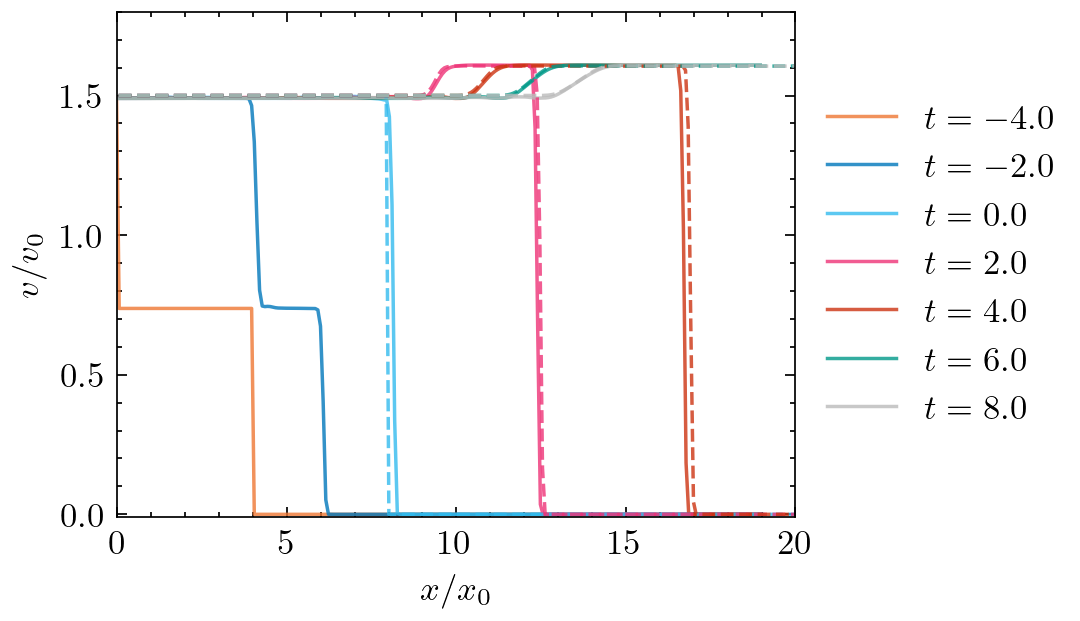}
    \caption{Athena++ simulation of colliding shocks with $v_{\mathrm{sh},1} = 1$, $v_{\mathrm{sh},2} = 2$ (solid line) and corresponding Sod shock tube (dashed line) with $v_\mathrm{R} = 1.5$,  $v_\mathrm{L} = 0$. The shocks collide at $x = 8$ at $t = 0$, which is when the Sod shock simulation starts.}
    \label{fig:sodshock}
\end{figure}

We consider two shocks with high speed that could be produced in the Galactic disk and move outwards. We set the first shock's velocity as $v_{\mathrm{sh},1} = 400\,\mathrm{km/s}$ and the second shock's velocity to be twice as large $v_{\mathrm{sh},2} = 2 v_{\mathrm{sh},1}= 800\,\mathrm{km/s}$ (note this value satisfies eq.~(\ref{eq:RHvs2})). Since the first shock moves into a low pressure medium it has a compression of $4$ and, thus, accelerates particles to a spectrum with $-2$ slope. The compression ratio of the second shock is only $q = 2.5$, so it produces a steeper spectrum $s = 2 - 3q/(q-1) = -3$ \citep{Drury}. 

By considering planar shocks, we solely investigate the impact of the colliding shocks on the energy spectrum, without any influence from the shock geometry or spherical expansion. 

\subsection{Diffusion: Bohm diffusion in constant magnetic field}

Like in the previous section we consider Bohm diffusion, $\kappa = 1/3 ( p \mathrm{c} / (q B) )$. Here, the magnetic field strength stays constant and $B = 1\,\mu\mathrm{G}$. Thus, the diffusion coefficient only scales with momentum and the drift term vanishes, $\partial \kappa / \partial z = 0$. 

In contrast to spherical shocks, for one-dimensional planar shocks, there are no geometric constraints for the confinement of particles. However, it is questionable if particles with energy higher than $10-100\,\mathrm{PeV}$ find waves to scatter with. If not, the ensemble averaged description is not applicable.

\subsection{Estimate of the maximal energy}

\begin{figure}
    \centering
    \includegraphics[width=1\linewidth]{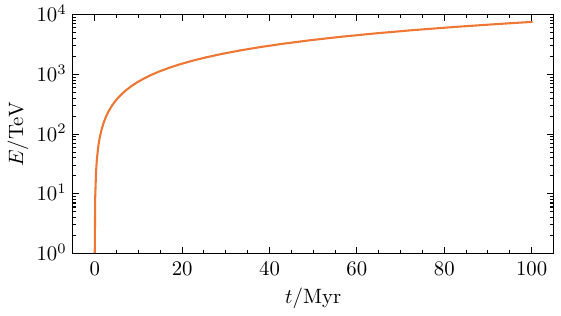}
    \caption{Estimate of the maximal energy that protons reach at a single
shock over time for $B_0 = 10\,\mu\mathrm{G}$, $v_\mathrm{sh} = 400\,\mathrm{km/s}$ and $E_0 = \mathrm{TeV}$ at $t_0 = 0$.}
    \label{fig:max_energy_planar}
\end{figure}

An estimate of the maximal energy reached at one individual shock over time is obtained from eq.~(\ref{eq:Lagage}) analogous to the previous section. Here, the shock speed does not change over time. Assuming Bohm diffusion, the maximal energy is 
\begin{equation}
\label{eq:max_energy_planar}
    E_\mathrm{{max}}(t) = \frac{3}{20} q \mathrm{c} B v_\mathrm{sh}^2 \left(t - t_0\right) + E_0.
\end{equation}
The maximal energy grows linear in time and is shown for protons in fig.~\ref{fig:max_energy_planar} for $B = 1\,\mu\mathrm{G}$, $v_\mathrm{sh} = 400\,\mathrm{km/s}$ and $E_0 = \mathrm{TeV}$.

\subsection{Estimate of the spectral slope}

It has been suggested that such a system of shocks produces broken power-laws since the compression ratio of the subsequent shocks is lower. However, when two shocks are catching up with each other, at some point particles may be able to encounter both shocks, when their diffusion coefficient is large enough (See also the work by e.g.\ \citet{Bykov-etal-2013}, \citet{Wang-etal-2019} and \citet{Vieu-etal-2020}) for similar considerations in steady-state and time-dependent colliding/converging shock waves.)\footnote{While writing up the manuscript we became aware of the work by \citet{Vieu-etal-2020}. Their considerations for converging shock waves are similar to ours for colliding shocks, although not the same. Nevertheless, the resulting energy spectra agree very well.}
. In that case, the spectrum at the shock(s) should harden, since the total compression ratio is higher\footnote{Another way to interpret this is that the escape probability is changed.}. Here, the total compression ratio would be $q = q_1 q_2 $. For our simulations with $v_{\mathrm{sh},2} = 2 v_{\mathrm{sh},1}$ , we have $q= 10$ and a corresponding spectral slope $s \approx -1.33$. However, there is only a finite time, namely the time until the shocks collide, to actually accelerate particles to that harder spectrum. 

\begin{figure}
    \centering
    \includegraphics[width=1\linewidth]{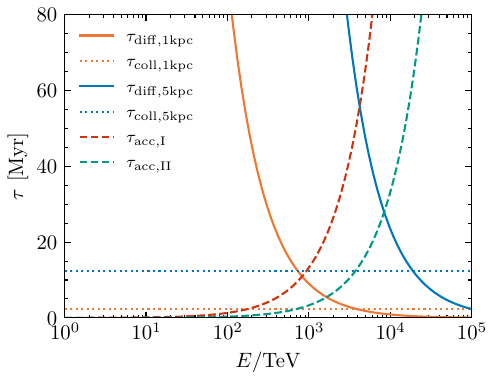}
    \caption{Time scales depending on the particles energy: Acceleration times at the respective shocks (dashed lines) increase with energy. The diffusive time scale (solid) for two different distances between the shocks $\Delta x = [1,5]\,\mathrm{kpc}$ decreases with energy. The collision time depending on $\Delta x $ is indicated by the dotted lines.}
    \label{fig:diff_timescale}
\end{figure}

For the spectrum to be accelerated up to the slope that corresponds to the total compression ratio, the diffusion time scale $\tau_\mathrm{diff} = \Delta x^2 / \kappa(p)$, where $\Delta x$ is the distance of the shocks, must be equal to the acceleration time scale $\tau_\mathrm{acc}$. Microscopically, each particles is then as likely accelerated at the first shock as it will encounter the second shock. When the diffusive time scale is longer but still on the order of the acceleration time, the hard spectrum cannot fully develop and the resulting spectral slope is somewhere between $s(q_1)$ and $s(q_1q_2)$.
Considering Bohm diffusion, fig.~\ref{fig:diff_timescale} shows the acceleration time scale of the first shock $\tau_\mathrm{acc, I}$ and second shock $\tau_\mathrm{acc, II}$ and the diffusion time scale for different distances $\Delta x$ between the two shocks depending on the particles momentum. Particles having a momentum on the order of the momentum defined by the crossing of the time scales or higher ($p>p_\mathrm{cross}$, with $\tau_\mathrm{acc}(p_\mathrm{cross}) = \tau_\mathrm{diff}(p_\mathrm{cross})$) are able to probe both shocks and are accelerated to a flatter (harder) spectrum than the one expected by an individual shock. The collision time is indicated by the dotted lines in fig.~\ref{fig:diff_timescale}. Note that, they might have an acceleration time scale that is higher than the time left until the shocks collide. 

In this scenario, we expect a flattening of the spectrum only for high energy particles $> \mathrm{PeV}$ shortly before the shocks collide: 1.) Particles need to be accelerated first to sufficiently high energy. The closer the shocks are, the lower is the critical energy to probe both shocks. 2.) The acceleration time scale to accelerate to the harder spectrum must be on the order of the collision time. This constraint is easier to fulfill at the faster, second shock.

\subsection{Simulation set-up}

\texttt{Candidates} are injected at $t = 0$ monoenergetically with $E = 1\,\mathrm{TeV}$ uniformly in the range $z = [0, 60]\,\mathrm{kpc}$. The first shock is launched at $t = 0$ with the speed $v_{\mathrm{sh},1} = 400\,\mathrm{km/s}$, the second shock emerges after $\Delta T$ with the speed $v_{\mathrm{sh},2} = 800\,\mathrm{km/s}$. The simulation ends at $t = 100\,\mathrm{Myr}$ and two time offsets between the shocks $\Delta T = [25, 40]$ are tested.

The diffusive step $\sqrt{\kappa(p) \Delta t}$, increases with energy. Here, the shockwidth is not scaled with energy, since we are interested in the interaction of the two shocks. Scaling the shock width with energy would cause different merging times as a function of pseudo-particle energy. Instead the time step $\Delta t$ is scaled linearly with the energy to keep the diffusive step within bounds and to fulfill eq.~(\ref{eq:constraints}). 

Again, the \texttt{CandidateSplitting} module is used to increase statistics at high energy. 

\subsection{Simulation results}

\begin{figure*}
    \centering
    \includegraphics[width=1\linewidth]{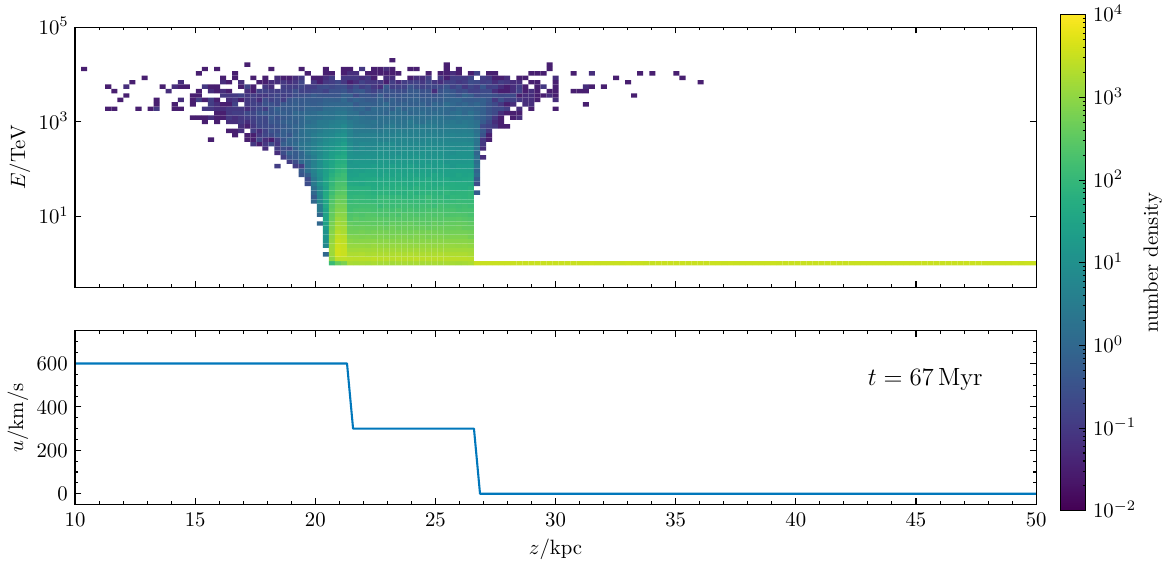}
    \includegraphics[width = 1\linewidth]{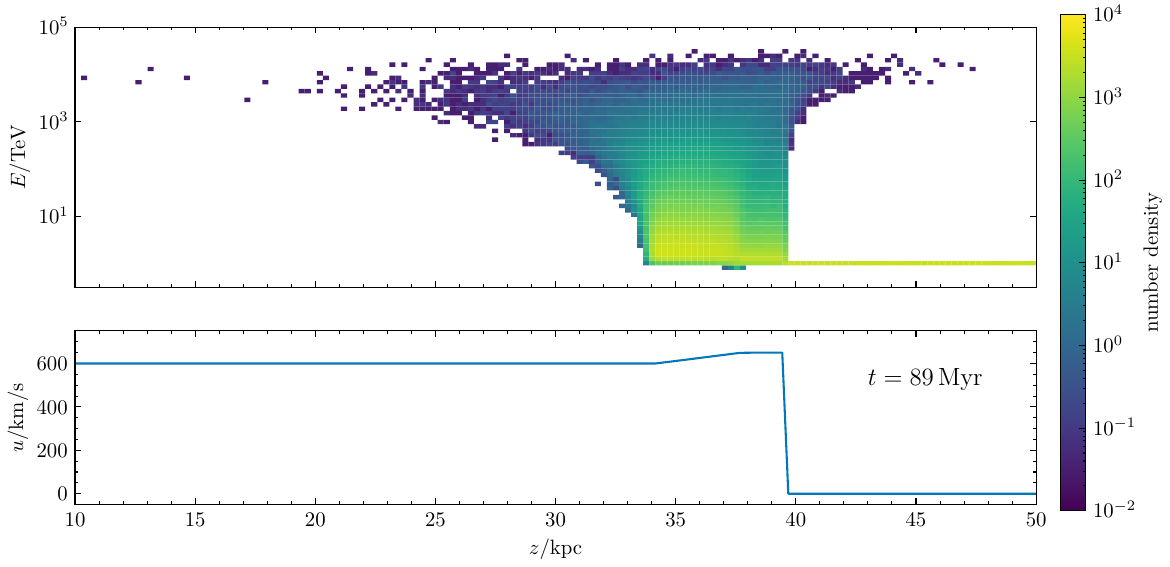}
    \caption{Wind profiles and space-energy histogram at $t = 67\,\mathrm{Myr}$ before the collision and $t = 89\,\mathrm{Myr}$. Particles are accelerated fast and the characteristic upstream and downstream spectra are visible for energy-dependent diffusion. }
    \label{fig:coll_hist}
\end{figure*}

\begin{figure}
    \centering
    \includegraphics[width=1\linewidth]{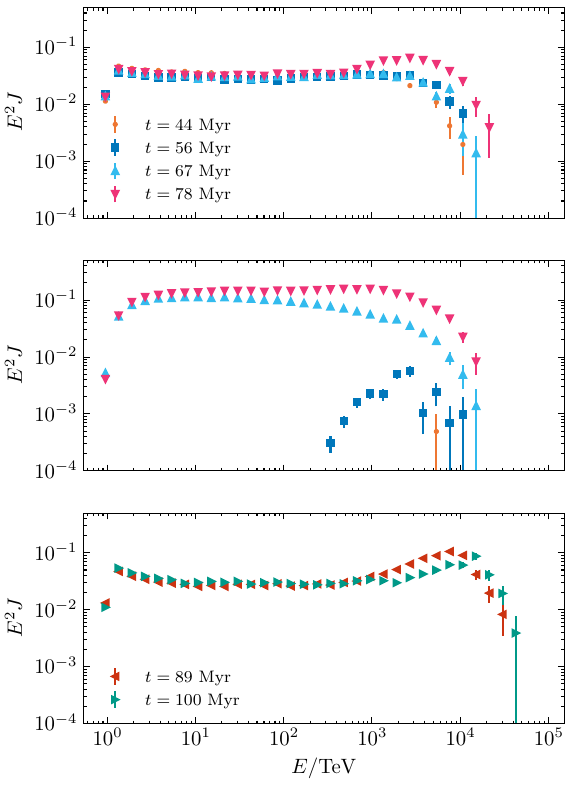}
    \caption{Energy spectra over time at the first shock (top), the second shock (middle) and the shock after collision (bottom). With $\Delta T = 40\,\mathrm{Myr}$ the shocks collide at $t = 80\,\mathrm{Myr}$. Around the collision, the spectrum flattens for $E > \mathrm{PeV}$ right before the energy cut-off.}
    \label{fig:coll_spectra_dT=40}
\end{figure}

Figure \ref{fig:coll_hist} shows the space-energy histogram at two times, before and after the collision of the shocks. In fig.~\ref{fig:coll_spectra_dT=40}, we show the spectra at the first shock, the second shock launched $40\,\mathrm{Myr}$ after the first, and the one that emerges after the collision at $t = 80\,\mathrm{Myr}$. 

Particles reach high energy fast, which is in agreement with eq.~(\ref{eq:max_energy_planar}). The first shock drags the particles along and at early times, only the high energy particles can be re-accelerated at the second shock (see e.g.~$t=56\,\mathrm{Myr}$ in fig.~\ref{fig:coll_spectra_dT=40}). At later times, the full downstream spectrum of the first shock reaches the second ($t=67\,\mathrm{Myr}$ one as can be seen in fig.~\ref{fig:coll_spectra_dT=40}). The second shock also pushes the particles back to the first one, which impacts in particular the low energy particles (see the upper panel of fig.~\ref{fig:coll_hist}). The flux $E^2 J$ is therefore higher at the second shock. At $t= 78\,\mathrm{Myr}$, shortly before the shocks collide, the spectrum at the second shock flattens. At the first shock, a flattening at high energies $E > \mathrm{PeV}$ right before the energy cut-off becomes visible. This is consistent with the findings by \citet{Vieu-etal-2020}. 

Once the shocks collide, the new shock emerges. It is faster than both of the previous ones and accelerates particles to even higher energy (see eq.~(\ref{eq:max_energy_planar})). Behind the shock a rarefaction wave develops and particles are cooled in that region. The flattening of the spectrum at high energies is still visible, however, the shock now accelerates particles to a $-2$ spectrum as it is again a strong shock. Since from the merger on only low energy particles $E_0 = \mathrm{TeV}$ are injected at the shock, we expect the spectrum to relax back to a $-2$ slope.

\begin{figure}
    \centering
    \includegraphics[width=1\linewidth]{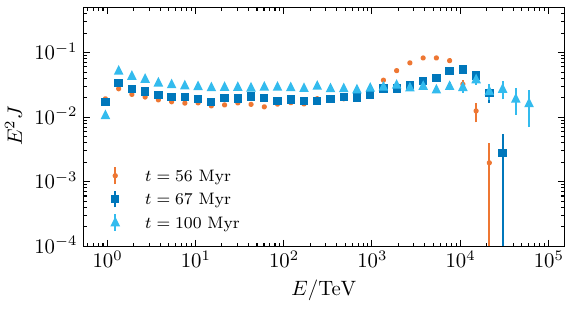}
    \caption{Energy spectra over time at the shock after collision. With $\Delta T = 25\,\mathrm{Myr}$ the shocks collide at $t = 50\,\mathrm{Myr}$. Over time, the new shocks accelerates particles to higher energy with $-2$ slope and the bump at high energy disappears. }
    \label{fig:spectra_coll_dT=25}
\end{figure}

Figure \ref{fig:spectra_coll_dT=25} shows the spectra at the single shock after a collision at $50\,\mathrm{Myr}$. The second shock emerged faster, $25\,\mathrm{Myr}$ after the first. At $t = 100\,\mathrm{Myr}$, the spectrum is a single power-law with $-2$ slope again. Thus, in this scenario, the flattening of the spectrum is a transient effect. 

\subsection{Propagation back to the Galaxy}
\label{sec:backpropagation}

Without cooling, only few particles of the highest energy make it back to the Galactic disk in fig.~\ref{fig:coll_hist}. Such wind profiles are only expected locally, in a magnetic flux tube and for a finite time. For an estimate we assume that the wind ceases after $100\,\mathrm{Myr}$. The re-accelerated particles can now diffuse back to Galaxy without going against the wind and the time evolution of the number density $n$ becomes
\begin{equation}
    \frac{\partial \mathcal{N}}{\partial t} = \kappa \frac{\partial^2 n}{\partial z^2} + S(z, t) \quad,
\end{equation}
where the diffusion coefficient remains spatially constant. The spectra that come back to $z = 5\,\mathrm{kpc}$ are obtained from the convolution of the particle distribution at $t=100\,\mathrm{Myr}$ with the Green's function of the pure diffusion equation
\begin{equation}
   \mathcal{N}(z,t) = \frac{1}{(4 \pi \kappa(p) t)^{1/2}} \mathrm{exp}\left( - \frac{(z - z_0)^2}{4\kappa(p_0)t}\right),
\end{equation}
where $z_0$ is the particles position at $t = 100\,\mathrm{Myr}$ and $p_0$ their momentum. Without the background wind, there are no processes to gain or loose energy. We test two diffusion scenarios: Bohm diffusion $\kappa = p\mathrm{c} / (q B)$ and Galactic diffusion $\kappa = 5\times10^{24}\mathrm{m^2/s}\,(E/\mathrm{GeV})^{1/3}$.

\begin{figure}
    \centering
    \includegraphics[width=1\linewidth]{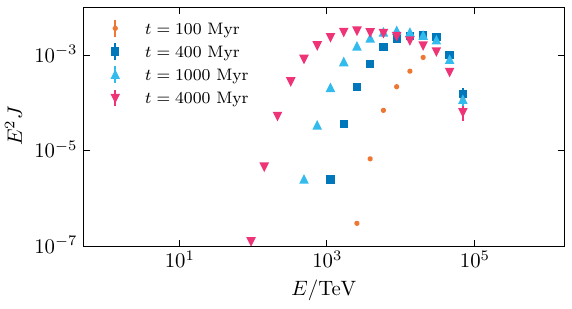}
    \includegraphics[width=1\linewidth]{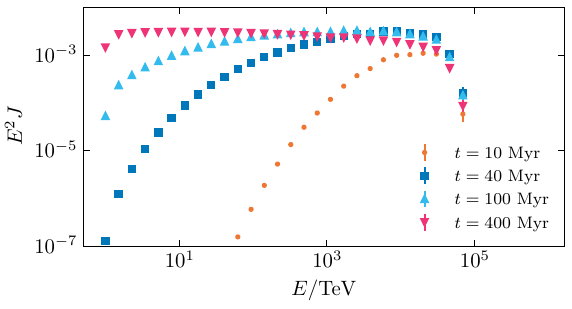}
    \caption{Spectra at $x = 5\,\mathrm{kpc}$ at differences times $t$ after the wind stopped at $t = 100\,\mathrm{Myr}$. The spectra are obtained from the convolution with the Green's function of the pure diffusion equation. Top: Bohm diffusion. Bottom: Galactic diffusion. }
    \label{fig:backprop}
\end{figure}

Figure \ref{fig:backprop} shows the particle spectra at $z = 5\,\mathrm{kpc}$ for different time durations after the wind has ceased. For Bohm diffusion, only particles re-accelerated to high energy make it back to the Galaxy, similar to the results obtained in the previous section for a spherical shock. At late times, more low energy particles $<\mathrm{PeV}$ contribute and the peak of the spectrum shifts to lower energy. The high energy particles $>\mathrm{PeV}$ spread over a wider distance and the probability to find them at $5\,\mathrm{kpc}$ decreases. The difference in the spectra between $\Delta T = 25\,\mathrm{Myr}$ and $\Delta T = 40\,\mathrm{Myr}$ between the two shocks is negligible.

The Galactic diffusion coefficient is higher and time scale at which high energy particles make it back is shorter. Also, $100\,\mathrm{Myr}$ after the wind has stopped a significant amount of particles with energy $<100\,\mathrm{TeV}$ arrives at $5\,\mathrm{kpc}$. 

\section{Discussion}
\label{sec:discussion}

We investigated scenarios that contribute to a dynamical shock landscape forming in the Galactic halo: Strong explosions, resulting from several SN or SB outbursts run out of the Galactic disk and experience a decreasing density profile. We find, that such shocks are strong enough to accelerate particles up to the spectral knee.  With SN or SB explosions happening frequently in the Galaxy, they will interact with each other \citep{Dorfi-etal-2019}. Thus, we studied the scenario of two shocks that catch up with each other and eventually collide. 
In both scenarios, we obtained the time-dependent spectra at the respective shocks and the spectrum of particles that is able to diffuse back to the Galaxy and contribute to the CR spectrum. We used a modified version of CRPropa that facilitates time-dependent wind profiles to simulate test particle transport and acceleration. 

In sec.~\ref{sec:implementation}, we used predictions for the maximal energy and spectral slope at a Sedov-Taylor blast wave to validate the new code. For spherical, expanding shocks the spectral slope can deviate from the well-known $-2$ slope obtained for planar shocks, due to the different injection rate at the shock (the shock surface grows with time) and adiabatic cooling in the expanding wind on the downstream side of the shock. The correction is negligible if the diffusion coefficient is small compared to the shock radius and shock speed, $\kappa \ll (R\dot{R})$. 

With energy-dependent diffusion, this effect may become important for high energy particles, leading to a steepening of the spectrum before the cut-off. In our scenario, where we consider Bohm diffusion in a decreasing magnetic field, $B \propto r^{-2}$, and CR are assumed to be pre-accelerated to energies up to $100\,\mathrm{TeV}$, we find that they are re-accelerated to a steeper spectrum $ s < -2$ for energies in the range $10-100\,\mathrm{PeV}$. Thus, the steepening of the spectrum between the knee and ankle could partially be a result of the shocks geometry and magnetic field turbulence.

To accelerate particles up to that energy, the shock speed needs to be high enough, which we only find for $E_\mathrm{blast} = 10-100\,E_\mathrm{SN}$ when the density decreases with distance to the Galaxy. Here, we used $\rho \propto r^{-2}$ to employ the critical solution found by \citet{Isenberg-1977}. This is reasonable when the outburst happens close to the edge of the Galactic disk or, when another explosion previously happened in the vicinity and swept the surrounding mass away. 

We find that, re-accelerated particles from $100\,\mathrm{TeV}$ to $100\,\mathrm{PeV}$, make it back to the Galaxy. Where $100\,\mathrm{PeV}$ are only reached when the magnetic field is strong/diffusion is small. Since the wind expands and slows down over time, it is easier for CR to propagate back than, e.g.\ in the scenario of a Galactic wind termination shock, where particles need to escape upstream and diffuse against the strong, unshocked wind (e.g.\ \citet{JokipiiMorfill87} and \citet{Merten-etal-2018}). Also, the distance they must cover is way smaller than in the case of a Galactic termination shock at $r \approx 200\,\mathrm{kpc}$. We even find that the flux from $100\,\mathrm{TeV}$ to $100\,\mathrm{PeV}$ is rather stationary and does not change much on time scales of $80\,\mathrm{Myr}$ after the explosion. This is due to the small diffusion and wind speed close to the Galaxy. Thus, only few strong explosions in the past are required to obtain a stationary flux in that energy range. 

The considered spherical geometry has, on the other hand, a disadvantage for the low energies: CR with small diffusion coefficients ($E < 100\,\mathrm{TeV}$) are pulled outwards with the wind and their flux decreases over time. However, in reality such outbursts are likely not spherically symmetric. The global wind is likely asymmetric, and the explosion will depend significantly on the local environment of the SB. A more realistic shock could probably re-accelerate CR to high energies $(\gtrsim \mathrm{PeV})$ without reducing the number of low energy CRs. 

It is interesting to further investigate the transport of CR that were re-accelerated at such local outbursts, allowing for a more detailed study on the transport regimes and the magnetic field geometry, which is beyond the scope of this work. The spectrum at the shock can be split up in two parts that are treated in a diffusive and ballistic way, respectively. The transition from diffusive to ballistic transport is already indicated, e.g.\ in fig.~\ref{fig:Sim1_hist}. Furthermore, the diffusion coefficient could increase to the Galactic value for CRs that escape the region close to the shock with amplified magnetic field. As shown in fig.~\ref{fig:backprop} it is more likely that they diffuse back.

Interestingly, the re-accelerated CR flux that comes back to the Galaxy is stationary on time scales $\lesssim 80\,\mathrm{Myr}$. Thus, in order to explain a stationary flux in that energy range, the strong outbursts do not necessarily have to happen frequently. A spiral magnetic field structure would further delay the CR arrival time \citep{Merten-etal-2018} and with that past explosions, when the Galaxy was more active in star formation (see e.g. \cite{Freundlich-2024} for the general time evolution of the star formation rate), could still contribute to the spectrum above the spectral knee.  

The amount and frequency of such explosions can be constrained by a transport study of CRs from various outbursts back to Earth. The source regions can be placed randomly in the Galaxy and in time (see \citet{Mertsch-2018} for a similar study). The source distribution can match the current structure of star forming regions \citep{Zhang-etal-2024}. These studies could constrain the anisotropy of arrival directions in the respective energy range. 

For shocks catching up with each other we showed, that in the one-dimensional planar case, the second shock has to be faster than the first shock. This is also true in a flux tube geometry, when the density behind the first shock decreases and was found from simulation by \citet{Dorfi-etal-2019}. We further showed, that the wind profile can be described by a modified version of the Sod's shock tube problem which has an analytic solution. Using results from Athena++ simulations of the colliding shocks, we constructed the wind profile and its derivative for use in CRPropa, allowing us to model particle transport and acceleration. The 1D model considered here sets the most stringent criterion possible for shocks to collide. Dropping the assumption would allow shocks with different launch footprints to collide obliquely which also produces a region of shear, which might become turbulent.

For the resulting energy spectrum there are two predictions: With the secondary shocks having a lower compression ratio, the energy spectrum of particles re-accelerated at those shocks steepens. And, when shocks are close to each other and particles encounter both shocks, the energy spectrum may become harder. 

In our scenario, we do not find a steepening of the energy spectrum at the second shock. First, only high energy particles reach the second shock and their acceleration time scale is too long to see the steepening of the spectrum (in contrast to the cut-off). Later, the acceleration process is constrained by the time left before the merger. However, taking cooling e.g.\ due to spherical expansion into account, the downstream spectrum of the first shock could reach the second shock more easily against the wind. Thus, it could be re-accelerate by the second, less compressed shock, likely to a steeper spectrum. 

We only find harder spectra close to the collision time similar to the analysis by \citet{Vieu-etal-2020} for converging flows. The diffusive time scale $\Delta x^2 / \kappa$ and the acceleration time scale $\propto \kappa$ must be on the same order which is only fulfilled when the shocks are already close. The spectrum relaxes back after the collision, when the newly formed shock accelerates particles again to a $-2$ spectrum. Nevertheless, we find the effect interesting, in particular in systems where the shocks can be assumed stationary and the acceleration time is not constrained by the collision time, e.g.\ \citep{Bykov-etal-2013, Malkov-Lemoine-2023}. Furthermore, the time-dependent test particle simulations with CRPropa can be extended to system with more complex geometry and/or diffusion tensors. 

In the case of 1D planar shocks, the wind speed remains high without radiative cooling and prevents particles from diffusing back to the Galaxy even if they escape downstream. Considering a spherical expansion particles are more likely to propagate back but are also subject to cooling.
However, considering a finite lifetime of the shock (here we used $100\,\mathrm{Myr}$ as in \cite{Merten-etal-2018} for the Galactic wind termination shock), particles have a higher chance to diffuse back to the Galaxy. We convolved the energy spectra at $100\,\mathrm{Myr}$ with the Green's function of the pure diffusion equation to find the time-dependent spectra at the Galactic boundary. As the analytic solution for the pure diffusion transport equation is known, no additional simulation has to be done for this. This has the advantage that different diffusion scenarios for the time after the wind ceased can be tested easily. 

Depending on the diffusion coefficient, only the particles re-accelerated to high energies $E > \mathrm{PeV}$ or the full spectrum $E > \mathrm{TeV}$ make it back to the boundary within $\approx 400\,\mathrm{Myr}$. Thus, a good description of the diffusion properties is crucial to obtain results that can be fitted to data. Considering a more realistic magnetic field structure, the arrival times at the Galactic boundary will change and likely become latitude dependent. Together with a study on the anisotropy of the diffusion coefficient also predictions for the anisotropy of the CR flux can be made (see e.g.~\citep{Merten-etal-2018}). 

In summary, we presented a modified version of CRPropa that facilitates time-dependent wind profiles and validated the new code by simulating particle acceleration at a Sedov-Taylor blast wave. We further applied this code to two acceleration scenarios: At a shock driven by an explosion that runs out of the Galaxy, and at two shocks catching up with each other and collide. In both cases, we find that protons can be accelerated up to $10-100\,\mathrm{PeV}$, depending on the wind speeds and magnetic field strength/turbulence. Since the shocks are closer to the Galaxy than a Galactic wind termination shock, particles are more likely to diffuse back to the Galactic boundary, either against the expanding wind or after the wind ceased. We further find that as the two shocks get close to one another, the spectrum hardens as particles encounter both shocks. 

The results can be applied also to other galaxies in the future. Even though the downstream escape is towards the galaxy, re-accelerated particles could escape the galaxy once the wind ceases --- similar to the study in sec.~\ref{sec:backpropagation}. Furthermore, the transport and acceleration of electrons can be modeled to predict the inverse Compton and synchrotron emission on shocks in the Galactic halo as predicted by \citet{Zhang-etal-2024}. 

\begin{acknowledgements}
SA, LM, and JBT acknowledge support from the Deutsche Forschungsgemeinschaft (DFG): this work was performed in the context of the DFG-funded Collaborative Research Center SFB1491 "Cosmic Interacting Matters - From Source to Signal" (grant no.~445990517). RH and EZ acknowledge support from NASA FINESST grant No. 80NSSC22K1749 and NSF grant AST-2007323 during the course of this work. This work was supported by the Ruhr University Research School funding line PR.INT (SA). This work has made use of the following software packages: Athena++ \citep{Athena++}, CRPropa \citep{CRPropa3.2}, matplotlib \citep{Hunter-2007}, numpy \citep{Harris-etal-2020}, pandas \citep{McKinney-2010}, and jupyter \citep{Kluyver-etal-2016}.

\end{acknowledgements}

\bibliography{references}
\bibliographystyle{aa}

\end{document}